\documentclass[useAMS,usenatbib]{mn2e}

\usepackage{graphicx}
\usepackage{txfonts}
\def \arcdeg {\hbox{$^\circ$}}                
\def \arcmin {\hbox{$^\prime$}}                
\def \arcsec {\hbox{$^{\prime\prime}$}}        
\def \msol  {\hbox{M$_{\odot}$}}               
\def \lsol  {\hbox{L$_{\odot}$}}               
\def \kms    {\hbox{${\rm km\,s}^{-1}$}}       
\def \perccc    {\hbox{${\rm cm}^{-3}$}}       

\title[Wide-field imaging of NGC\,2264]{The structure of molecular gas associated with NGC\,2264:  wide-field $^{12}$CO and H$_2$ imaging}
\author[J. V. Buckle, J. S. Richer and C. J. Davis]{J. V. Buckle$^{1,2}$\thanks{E-mail:
j.buckle@mrao.cam.ac.uk}, J. S.
Richer$^{1,2}$ and C. J. Davis$^3$\\
$^{1}$Cavendish Astrophysics Group, J J Thomson Avenue, Cambridge, UK\\
$^{2}$Kavli InstituteÊfor Cosmology Cambridge, Madingley Road, Cambridge, UK\\
$^3$ Joint Astronomy Centre,	University	Park, Hilo, Hawaii, USA}
	
\begin{document}

\date{Accepted . Received ; in original form}

\pagerange{\pageref{firstpage}--\pageref{lastpage}} \pubyear{2010}

\maketitle

\label{firstpage}

\begin{abstract}

We present wide-field, high-resolution imaging observations in $^{12}$CO 3$\rightarrow$2 and H$_2$ 1--0\,S(1) towards a $\sim$1 square degree region of NGC\,2264. We identify 46 H$_2$ emission objects, of which 35 are new discoveries.  We characterize several cores as protostellar, reducing the previously observed ratio of prestellar/protostellar cores in the NGC\,2264 clusters. The length of H$_2$ jets increases the previously reported spatial extent of the clusters.  In each cluster, $<$0.5\% of cloud material has been perturbed by outflow activity. A principal component analysis of the $^{12}$CO data suggests that turbulence is driven on scale{\rm s $>$2.6~pc, which is} larger than the extent of the outflows. We obtain an exponent $\alpha$=0.74 for the size-linewidth relation, possibly due to the high surface density of NGC\,2264. In this very active, mixed-mass star forming region, our observations suggest that protostellar outflow activity is not injecting energy and momentum on a large enough scale to be the dominant source of turbulence. 
\end{abstract}


\section{Introduction}

Studies of young clusters are essential for understanding star formation, since most stars, and especially massive stars, are known to form in clusters \citep{zinnecker}. NGC\,2264 is an attractive target for studying this mode of star formation, since it is nearby, has relatively low foreground extinction and there is a large population of pre-main sequence stars, whose age spread is evidence of sequential star formation \citep*{adams}. 
The region contains a wealth of submm cores \citep{wardthompson}, pre-main sequence stars and protostellar sources \citep{young,teixeira2006}, many of which have formed in clusters \citep{wolfchase,furesz,young}.  At 760--900~pc distance \citep[see][and references therein for discussion]{baxter}, the region contains over 300 Class II objects, and 30 Class I objects \citep{dahm}. Several young Class 0 objects have been identified \citep*{teixeira2006,furesz,peretto2006}.  Kinematic motions include large-scale infall motions \citep{peretto2006,williamsgarland}, and large, fast outflows, including the well-known NGC\,2264~G \citep{lada,hedden}. 
The wide field optical images of \citet{reipurth} reveal a number of HH objects, several of which trace giant, parsec-scale outflows.  Published infrared images of the NGC\,2264 region cover only very small areas; maps of the NGC\,2264~G molecular outflow (MHO~1358/1359\footnote{The MHO (Molecular Hydrogen emission-line Object) catalogue is maintained at http://www.jach.hawaii.edu/UKIRT/MHCat/}) have been presented by \citet{davis1995}, while more recently a number of H$_2$ features (MHO~1349-1359) have been identified around NGC\,2264~C by \citet{wang}.

Outflows have a significant impact on material within the parent molecular cloud, injecting energy and momentum at large distances. The outflows extend typically over 0.1--1 parsec, and the combined sub-arcsecond resolution H$_2$ images with deep spectral imaging in $^{12}$CO,  result in a more complete census of star formation activity in clustered regions. Outflow kinetic energies can be comparable to, or even larger than, turbulent and gravitational energies of natal clouds, and could have a significant disruptive effect \citep{mckee2007}. In order to find evidence for, and quantify the disruptive effects of molecular outflows, it is crucial to map the entire cloud. 

   \begin{figure}
   \centering
   \includegraphics[width=8cm,angle=0]{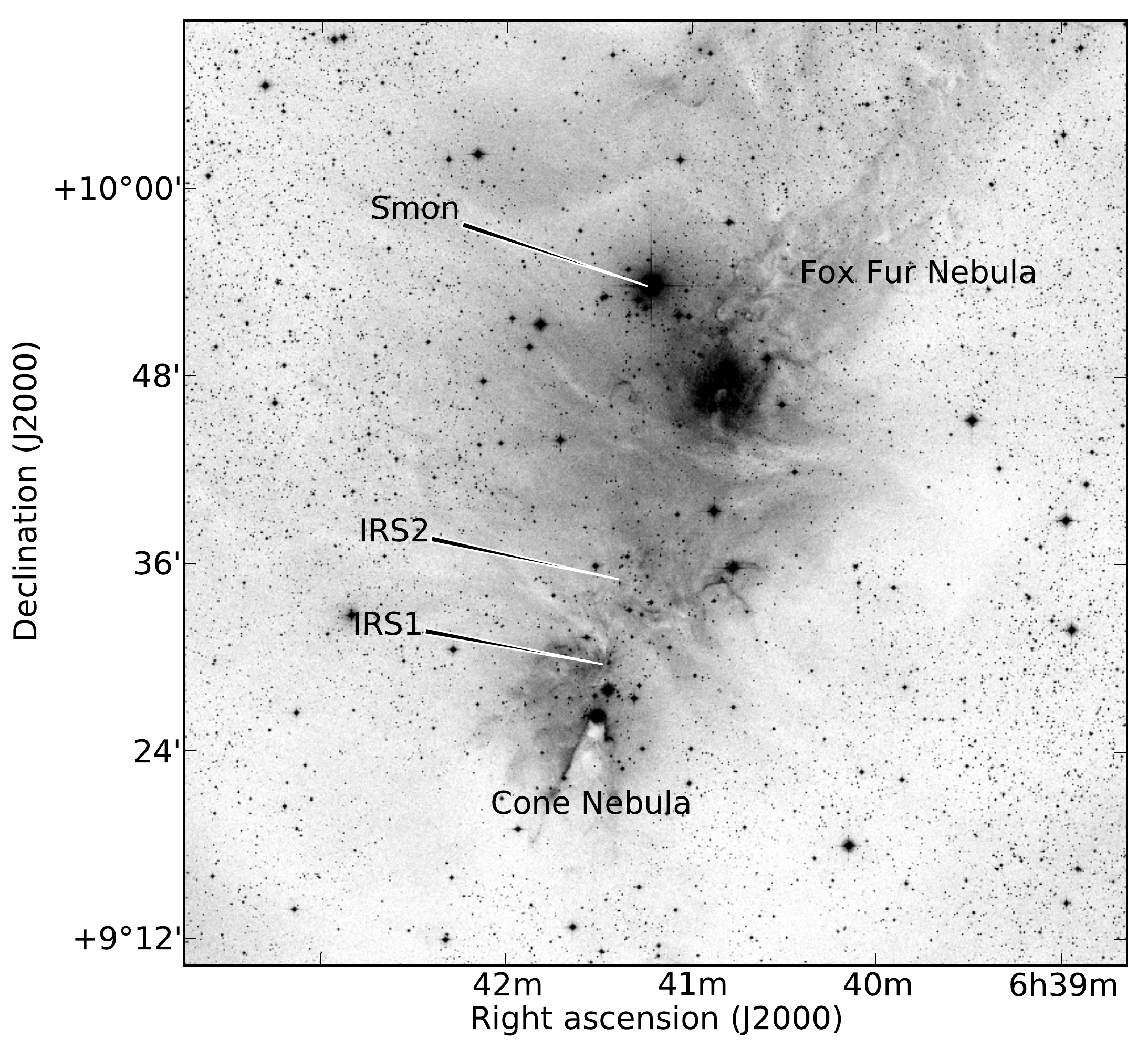}
  \caption{DSS optical image of NGC\,2264 (the Christmas Tree Nebula) showing the massive O star S\,Mon, the two star-forming clusters NGC\,2264~C (also known as Allen's Source, and IRS1) and NGC\,2264~D (also known as IRS2), and the two nebulae within the cloud, the Cone Nebula and the Fox Fur Nebula.}
              \label{fig-optical}
    \end{figure}

Fig. \ref{fig-optical} shows the Digitized Sky survey (DSS) optical image of NGC\,2264, with the main star formation regions and nebulae labelled, obtained with Starlink Gaia using Skycat\footnote{http://archive.eso.org/cms/tools-documentation/skycat}. A detailed review of this region has been published by \citet{dahmb}.
We present wide field, high resolution mapping observations of NGC\,2264. Through CO observations we can provide a complete census of outflow activity in the region, investigating any dynamic impact the outflows are having on the parent molecular cloud through energy injection. Using H$_2$ narrow line imaging, we can detect the youngest flows, from emission arising in gas shocked by the impact of protostellar outflows, and measure the extent of these flows by tracing the detected emission knots and HH objects. 
\section{Description of Observations}

\subsection{Spectral Line Observations}

The spectral-line observations were taken at the JCMT (James Clerk Maxwell Telescope) using HARP \citep[Heterodyne Array Receiver Programme,][]{buckle}. Observations of CO 3 $\to$ 2 at 345.796~GHz were taken with a 0.488MHz (0.423 \kms) spectral resolution and gridded with 6$\arcsec$ spatial pixels.
The beam size is 14$\arcsec$ at 345~GHz. The map covers an area of 1 square degree, and observations totalling 12.9 hours were taken during August and October 2007, in position-switch raster mode. The observations were taken in Band 3 weather, with zenith opacity values between 0.13 and 0.19, and average system temperatures $\sim$550~K. Averaged over emission free regions of the map, a 1$\sigma$ rms of 0.32~K per pixel in 1.0~\kms\ spectral channels was obtained.

All intensities reported here are in units of T$^*_{\rm A}$, the antenna temperature corrected for sky and telescopes losses \citep{kutner}. The main beam brightness temperature, T$_{\rm mb}$ = ${\rm T^*_{\rm A}}/\eta_{\rm mb}$, with a main beam efficiency, $\eta_{\rm mb}$~=~0.66. Frequent pointing and focus observations were carried out, and calibration observations using N2071IR. The scatter in these observations suggest a calibration accuracy of $\sim$20\%.

The data were reduced and analyzed using the Starlink project
software, in particular {\sc SMURF} \citep{jenness} and {\sc KAPPA} \citep{currie} routines.

\subsection{H$_2$ Imaging Observations}

The infrared imaging observations were taken at UKIRT (the United Kingdom Infrared Telescope) using the near-IR wide-field camera WFCAM \citep{casali}.  The observations cover an area on sky of ~0.75 sq. degrees (a WFCAM `tile'), with a pixel size of 0.2$\arcsec$. Images were obtained through broad-band K ($\lambda = 2.20$~$\umu$m, $\delta\lambda = 0.34$~$\umu$m) and narrow-band H$_2$ ($\lambda = 2.121$~$\umu$m, $\delta\lambda = 0.021$~$\umu$m) filters.  Exposure times of 5 sec and 40 sec were used; thus, the total per-pixel integration times were 100 sec and 800 sec in K and H$_2$, respectively.
    
Data were initially reduced by the Cambridge Astronomical Survey Unit (CASU), and distributed through the Wide Field Astronomy Unit (WFAU) archive. Further data reduction and analysis was carried out using Starlink {\sc KAPPA} routines \citep{davis2009}. 

\subsection{Data Summary}

The two datasets presented here probe the most active regions of current star formation. CO $3\rightarrow2$ emission traces dense, warm gas typical in regions of star formation, with temperatures 10--50~K and gas densities of 10$^4$--10$^5$~\perccc. The higher critical density of the $3\rightarrow2$ line over lower excitation lines makes this transition a useful tool for tracing dense, more collimated flows from the youngest sources. The rovibrational H$_2$ 1-0~S(1) line at 2.122~$\umu$m arises in regions with temperatures $\sim$2000--4000~K, and densities 10$^3$--10$^5$~\perccc.  Emission from this line indicates the presence of deeply embedded driving sources, and traces the youngest outflows.

\label{obs-summ}

   \begin{figure}
   \centering
   \includegraphics[width=9cm,angle=0]{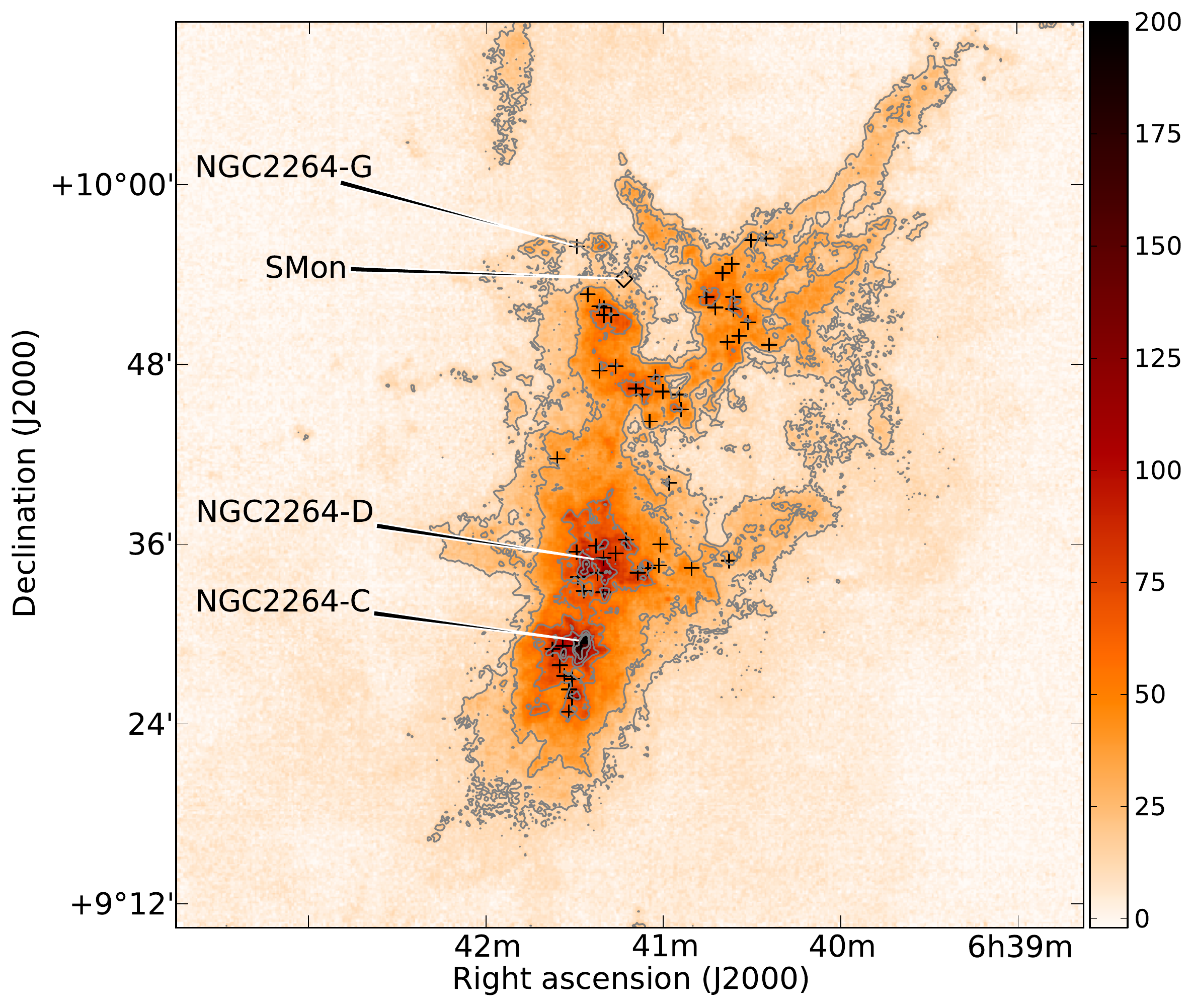}
  \caption{Integrated CO intensity map of NGC\,2264, in the velocity range -20 to 30 \kms, with contours at 15,30,60,90,120,150~K\,\kms.  A diamond marks the position of S\,Mon.  Crosses mark dense cores identified through SCUBA observations \citep{difrancesco}. Labels mark the well-known protostar NGC\,2264~G, and the two star-forming clusters NGC\,2264~C and NGC\,2264~D. }
              \label{fig-wco}
    \end{figure}

Fig. \ref{fig-wco} shows the integrated intensity map of CO $3\rightarrow2$ emission, in the velocity range -20 to 30 \kms (note all velocities are LSR unless otherwise stated).  The brightest integrated emission (246~K\,\kms) is towards the cluster NGC\,2264~C, while the brightest peak intensity (28~K) is towards the head of the Cone Nebula. Filaments can be seen extending south from NGC\,2264~C, east and west from NGC\,2264~D and north-west from the S\,Mon. S\,Mon is a massive (17.8~\msol) young star, of spectral type 07V \citep{herrero}, and the radiation pressure and wind from this star is likely to be heating and dynamically affecting nearby regions, such as the protostellar outflow source NGC\,2264~G \citep{teixeira2008,lada}. The two outflow lobes of NGC\,2264~G can be seen to the north-east of S\,Mon. West of S\,Mon, and extending southwards, is a region of very weak emission, which we refer to as a cavity, surrounded by a ring or bubble of emission containing several SCUBA dust cores. 

Dust pillars and cometary clouds, harbouring YSOs, are prevalent in {\it Spitzer} imaging of massive star forming regions \citep{smith2010}, where the emission is dominated by strong PAH features. The Cone Nebula resembles one of these cometary regions in CO emission. The structure of this emission suggests that molecular gas is being ablated, possibly by the action of the NGC\,2264~C cluster, exposing star forming cores. Red- and blue-shifted $^{12}$CO emission and H$_2$ emission at the head of the Cone indicates on-going star formation activity.

   \begin{figure*}
   \centering
  \includegraphics[width=18cm,angle=-90]{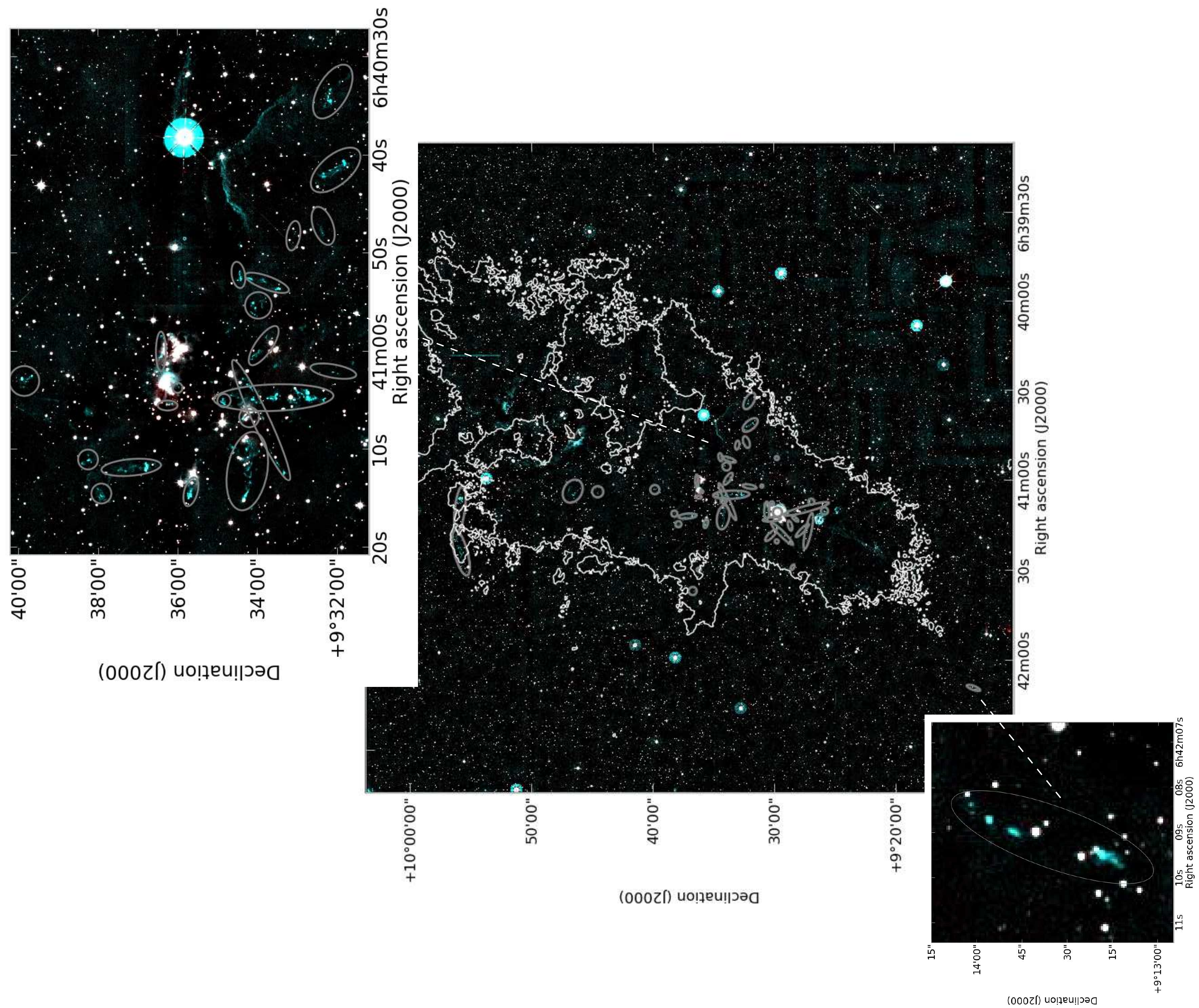}
  \caption{Colour composite image of the H$_2$ + continuum (blue and green) and K-band (red) data sets. Narrow-band H$_2$ line emission is seen in turquoise. The contour shows integrated CO emission at 15~K\,\kms. Ellipses mark the positions of MHOs we have detected, described further in the Appendix. Insets show zoomed images of two regions of newly detected MHOs, NGC\,2264~D and a region in the south-east. }
              \label{fig-h2}
    \end{figure*}

Fig.~\ref{fig-h2} shows a colour composite image of the H$_2$ 1-0~S(1) and K-band images of NGC\,2264\footnote{K- and H$_2$-band FITS data available from http://www.jach.hawaii.edu/UKIRT/TAP/}.  Ellipses mark the Molecular Hydrogen emission-line Objects (MHOs) we have detected. Inset images show zoomed versions of these data towards two regions, the cluster NGC\,2264~D and an isolated outflow to the south-east. H$_2$ line emission appears as turquoise in these images, which are shown in detail in Appendix Figs.~\ref{h2g} to~\ref{fig-outflow3}. We detect 46 MHOs, of which 35 are new discoveries, and described in detail in the Appendix. We detect regions of multiple jets and bow shocks surrounding NGC\,2264~C \citep[extending further than seen by][]{wang}, NGC\,2264~D and NGC\,2264~G \citep[as seen by][]{reipurth}. There are additional regions of jet activity to the east of NGC\,2264~D, and in the south-east corner of the map. Towards the Fox Fur Nebula and the Cone Nebula, we see much larger-scale arcs and filaments in H$_2$ emission.

\begin{figure*}
   \centering
\includegraphics[width=16cm,angle=0]{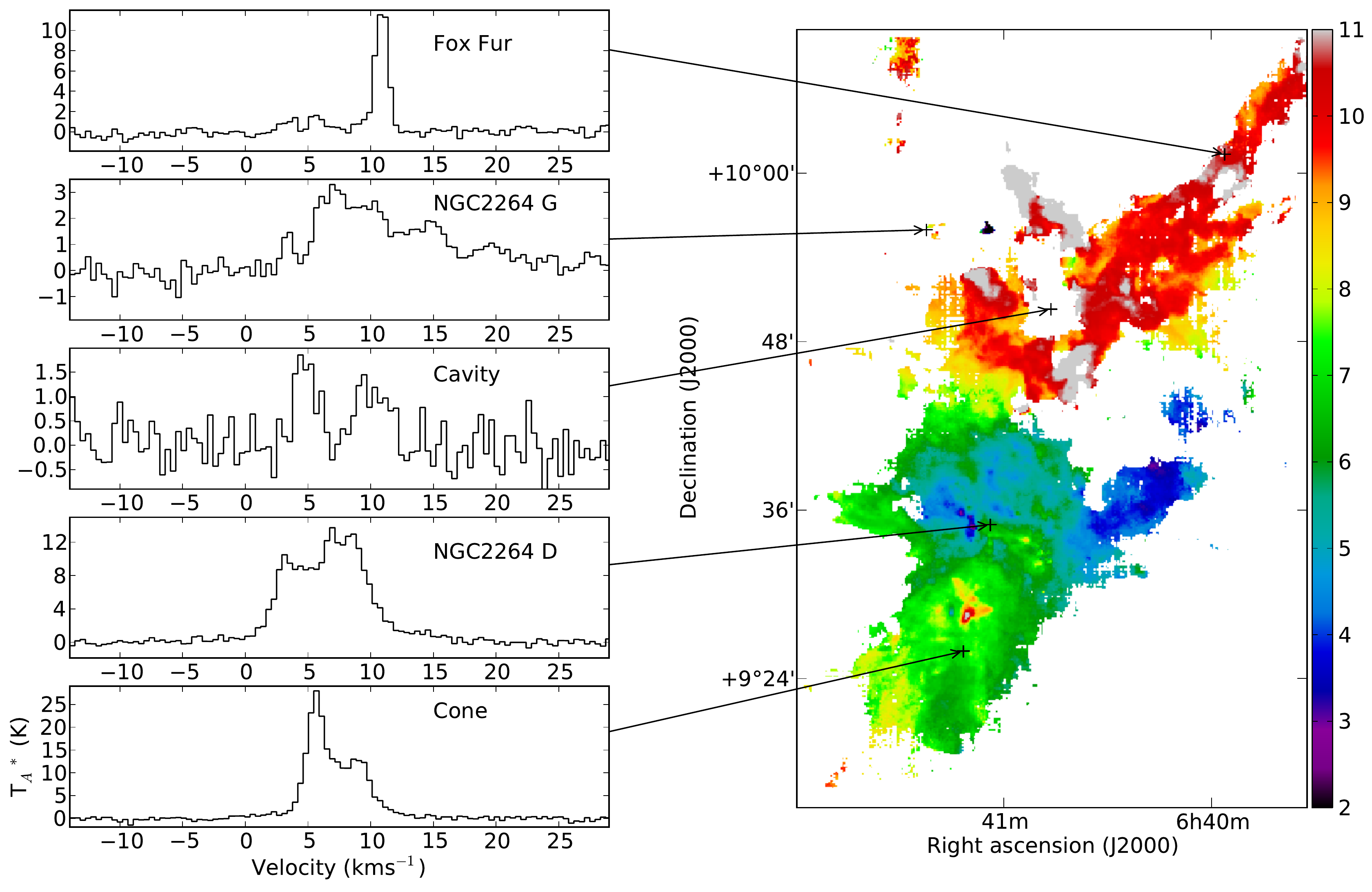}
  \caption{Spectra from individual pixels, from locations shown in the intensity-weighted velocity image, $\sum T_iv_i / \Delta v \sum T_i$, where $T_i,v_i,\Delta v_i$ are the intensity, velocity and velocity channel width respectively for each velocity channel, which is used to determine the velocity field. The intensity-weighted velocity data are thresholded to the $>$3$\sigma$ detection limit. The colour scale indicates the velocity field in units of \kms. }
              \label{fig-avspec}
    \end{figure*}

Representative CO spectra, and the intensity-weighted velocity map are shown in Fig.~\ref{fig-avspec}, highlighting the varying and complex velocity structure of this region. To the north, emission from the Fox Fur Nebula has narrow linewidths, and velocities $\geq$9~\kms. The star forming clusters, and NGC\,2264~G have extended line wings due to outflow activity. The clustered regions of star formation are kinematically distinct, with different cloud velocities, as has been previously noted \citep{maury,peretto2006}. The Cone region, and NGC\,2264~C, have velocities $\sim$7~\kms, while NGC\,2264~D has velocities $\sim$5~\kms. Several regions show multiple velocity components, including the cavity south-east of S\,Mon, the Cone Nebula and NGC\,2264~D.

\citet*{crutcher} and \citet{furesz} identify distinct regions across NGC\,2264 which are broadly in agreement with the velocity gradient and extent of the red, green and blue velocity regions in Fig.~\ref{fig-avspec} respectively. \citet*{sung2009} identifies and assigns ages to distinct star forming clusters using {\it Spitzer} data. The region surrounding S\,Mon, including NGC\,2264~G is aged at 3.1~Myr. The Cone Nebula is at a younger age, due to the number of embedded sources. \citet*{peretto2007} calculates very young ages for the clusters NGC\,2264~C and NGC\,2264~D of $\sim$0.1~Myr. We look at the regions of active star formation in more detail in the following section.

\section{Active Star Forming Regions}
\label{outflows}

The wide-field $^{12}$CO $3\rightarrow2$ and H$_2$ 1-0~S(1) data can be used to provide a comprehensive quantitative analysis of the star formation activity across NGC\,2264. Outflow and jet properties are expected to change as protostars evolve from the early accretion phase. The youngest objects, which are still accreting most of their final mass, are expected to have more powerful outflows, as measured from the CO momentum flux or jet luminosity \citep{caratti,arce,bontemps}.
In the following analysis, we use CO $3\rightarrow2$ to estimate the total outflowing mass, and the energy and momentum injected into the ambient cloud. We use H$_2$ 1-0~S(1) to estimate the total H$_2$ luminosity in the jets, and  place constraints on the impact of star formation activity on the ambient cloud.

For the emission in the outflow wings, we assume optically thin conditions, and calculate the total mass and energetics of molecular gas using emission from the $^{12}$CO $3\rightarrow2$ transition following \citet{garden1991,buckle2010}. \citet{maury} examined the opacity in the line wings of $^{12}$CO using CO isotopologues towards the NGC\,2264~C cluster, and found the emission to be optically thin outside of the central $\sim$10~\kms\ of the line profile. For the outflow contribution at low velocities, the optically thin assumption may not be correct, and the masses are then lower limits. Since we make no correction for the inclination angle of the outflows, the values for momentum and kinetic energy are lower limits. We use an excitation temperature of $T_{\rm ex}$~=~20~K in order to aid comparisons with previously published work \citep[e.g. ][]{lada}. The data are thresholded at the 3$\sigma$ level. 

Due to the NGC\,2264 velocity field (shown in Fig.~\ref{fig-avspec}), care needs to be taken in assigning velocity limits for the high velocity emission. We have taken the values from \citet{peretto2006}, who describe N$_2$H$^+$ observations, giving systemic velocities of 7.5~\kms\ for NGC\,2264~C and 5.5~\kms\ for NGC\,2264~D.  For the high velocity material, we use the velocity intervals $\pm$5.0~\kms\ to $\pm$20~\kms\ either side of the systemic velocity for each cluster. For other regions, the sources are sufficiently isolated for an estimate of the systemic velocity to be made from $^{12}$CO emission in areas adjacent to the molecular outflows. For NGC\,2264~G, we use a systemic velocity of 5.0~\kms, with ranges up to $\pm$30~\kms\ from the systemic velocity. Emission from ambient material covers the red lobe at low velocities, and so we use a lower velocity of 2~\kms\ for the red lobe, and --1~\kms\ for the blue lobe. For the protostar in the far south, IRAS2, we adopt a systemic velocity of 7.8~\kms, and velocity intervals of 2 to 10~\kms\ for the red-shifted emission, and --6 to --1~\kms\ for the blue-shifted emission.

We have extracted H$_2$ line emission fluxes from the H$_2$ plus continuum data after carrying out aperture photometry with background subtraction to calculate the calibrated integrated line flux. The H$_2$ plus continuum data is used rather than the continuum-subtracted data, since it does not add to the calibration uncertainties, and there are no deep negative components from an imperfect subtraction routine. Calibration was checked on stars in the field with known K-band magnitude, and is accurate at the 10\% level.

The quadrupole transitions of H$_2$ (known as O, Q and S branches) are the main cooling mechanism for shock-excited jets detected in the near-IR. The rovibrational transitions arising in the near-IR, including the H$_2$ 1-0~S(1) transition, can be used to estimate the total molecular hydrogen luminosity ($L_{H_2}$) of the flow. Younger flows are expected to have more powerful outflows, and higher values of $L_{H_2}$.

The total H$_2$ luminosity ($L_{H_2}$) can be calculated from the H$_2$ 1-0~S(1) luminosity ($L_{H_{2.12}}$), with $L_{H_2} \sim 10 \times L_{H_{2.12}}$ \citep{caratti}, for typical H$_2$ jet temperatures between 1500 and 2500~K. If emission from the jet arises in regions at higher temperatures, this method will underestimate $L_{H_2}$ by factors up to 2.5. $L_{H_2}$ will also be underestimated by up to an order of magnitude in regions of high extinction (A$_{\rm v}$=25).  \citet{caratti} have used near-IR spectral line observations to calculate the temperature and extinction towards one of the flows in our dataset, NGC\,2264~G. Their values of T=2100--2800~K, and A$_{\rm v}$=3--8~mag suggest $L_{H_2} = 10 \times L_{H_{2.12}}$ is a reliable estimate of $L_{H_2}$ for this region.

Table \ref{tab-mass} shows the mass, momentum and kinetic energies calculated for the red- and blue-shifted CO $3\rightarrow2$ emission, and the H$_2$ luminosities, towards each region. Individual MHOs included for each region are described in the Appendix. For the whole cloud, $L_{H_2}$=1.1~\lsol, although this is a lower limit, since we have not taken into account extinction. The masses are lower limits, since they are dependent upon temperature and opacity. If the high velocity emission is excited at temperatures $>$20~K, or the assumption of optically thin emission is not valid, then the masses will be increased. Additionally, it is not possible to entirely separate emission from outflowing gas from that of the ambient cloud. 

Towards all of the active star forming regions, we find more mass, momentum and energy in the red-shifted material than in the blue-shifted material.  \citet{maury} has examined the individual flows in NGC\,2264~C using CO and isotopic data, and also finds the red-shifted material to contain more mass and momentum. Asymmetry in molecular outflows is relatively common in low mass protostars, although no mechanism has  so far been proposed for the brightening of the red-shifted lobe.

\begin{table}
\begin{tabular}{@{}lrrrrrr}
\hline
&&\multicolumn{4}{c}{CO $3\rightarrow2$}&H$_2$\\
&    &Mass& Momentum  & Energy & Area& $L_{H_2}$\\
&     & &   & $\times$10$^{36}$&&\\
&     &\msol& \msol~\kms  & J& pc$^2$&\lsol\\
\hline
     NGC\,2264~C &   red &     1.70 &  14.6 & 146.0  &  1.7 & \\
       &  blue &   1.38 &  11.1 &  105.3  &  1.9 & \\
       & all &        &       &         &      &0.41$^a$\\
      NGC\,2264~D &   red &     2.63 &  21.8 & 212.9  &  3.4&  \\
       &  blue &   1.72 & 15.3 & 159.0  &  2.5 & \\
       & all &        &       &         &      &0.37\\
      NGC\,2264~G &   red &    0.55 &   7.1 & 111.2  &  0.8&  \\
       &  blue &   0.36 &  4.1 &  61.1  &  0.4 & \\
       & all &        &       &         &      &0.25\\
       IRAS\,2 &   red &     0.08 &   0.3 &   1.4  &  0.3 & \\
        &  blue &    0.04 &  0.1 &   0.3  &  0.4  &\\
       & all &        &       &         &      &0.01\\
\hline
\end{tabular}
$^a$ Emission from MHO1349 is excluded, since it can't be separated from the saturated continuum towards Allen's source.
\caption{Outflow parameters for the NGC\,2264 high velocity regions. }
\label{tab-mass} 
\end{table}

\subsection{The NGC\,2264 protoclusters}

\begin{figure*}
   \centering
   \includegraphics[width=14cm,angle=0]{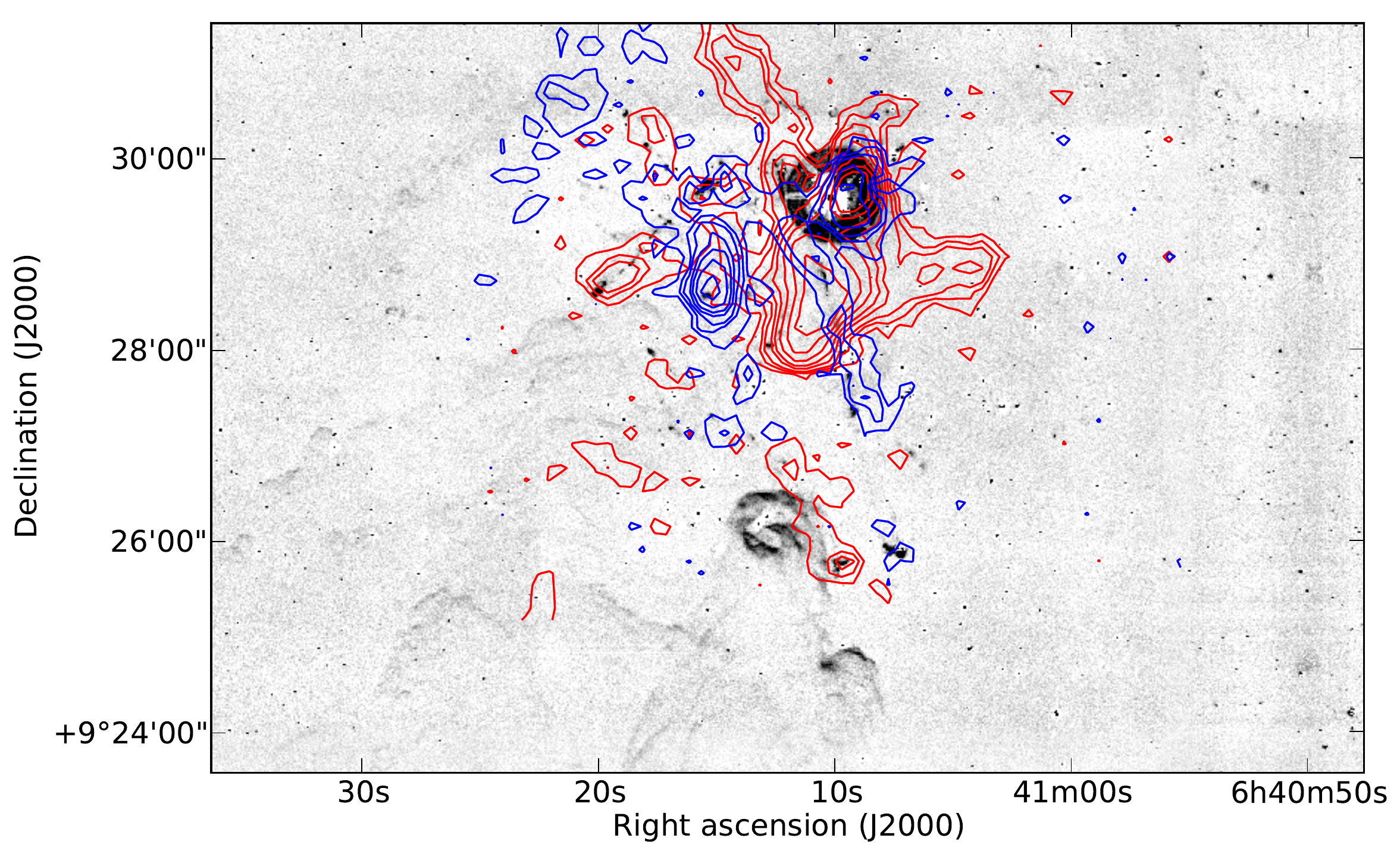}
  \caption{Red and blue shifted CO emission contoured over the H$_2$ continuum-subtracted greyscale image towards the NGC\,2264~C cluster. Contours are at 3.9, 6.5, 9.1, 14.3, 19.5, 27.3, 37.7, 48.1~K\,\kms.  Red-shifted emission extends from 12.2 to 30.9 \kms, and the blue-shifted from -17.8 to 2.1 \kms. The saturated source seen in the H$_2$ emission is Allen's source \citep{allen}, IRS1.}
              \label{fig-outflow2_main}
    \end{figure*}

NGC\,2264~D is a more massive cluster than NGC\,2264~C, with more protostellar sources, more mass and energy, and covering a larger area. Within the uncertainties, both clusters have similar H$_2$ luminosities. 
Fig.~\ref{fig-outflow2_main} shows the red- and blue-shifted CO emission contoured over the H$_2$ image towards NGC\,2264~C. The jets and outflows in the cluster can be seen extending past the Cone Nebula to the south. To the south-west of IRS1 is a red-shifted flow that contains the most high velocity gas, with line wings detectable out to $\sim$30~\kms. We have described in detail the individual H$_2$ objects and CO flows in the Appendix.

 \citet{peretto2006} identify 4 of the 12 NGC\,2264~C cores as prestellar. As described in the Appendix, all four of these cores are spatially located within collimated red- and blue-shifted CO emission, and compact knots of H$_2$ emission. Additionally, 3 of the cores are coincident with {\it Spitzer} 24~$\umu$m sources. These data suggest that these 4 cores are young Class 0/I protostars. \citet{sung2009} classifies 17 24~$\umu$m sources as protostellar in NGC\,2264~C.

\citet{peretto2006} estimate the total gas mass in NGC\,2264~C to be 1650~\msol, compared to the 3~\msol\ we estimate to be entrained in the outflow. Although there are multiple energetic outflows in this cluster, only $\sim$0.2\% of the molecular gas has been entrained by outflows, or affected by star formation activity. This is a lower limit, since it is dependent upon the opacity. 

\begin{figure*}
   \centering
     \includegraphics[width=14cm,angle=0]{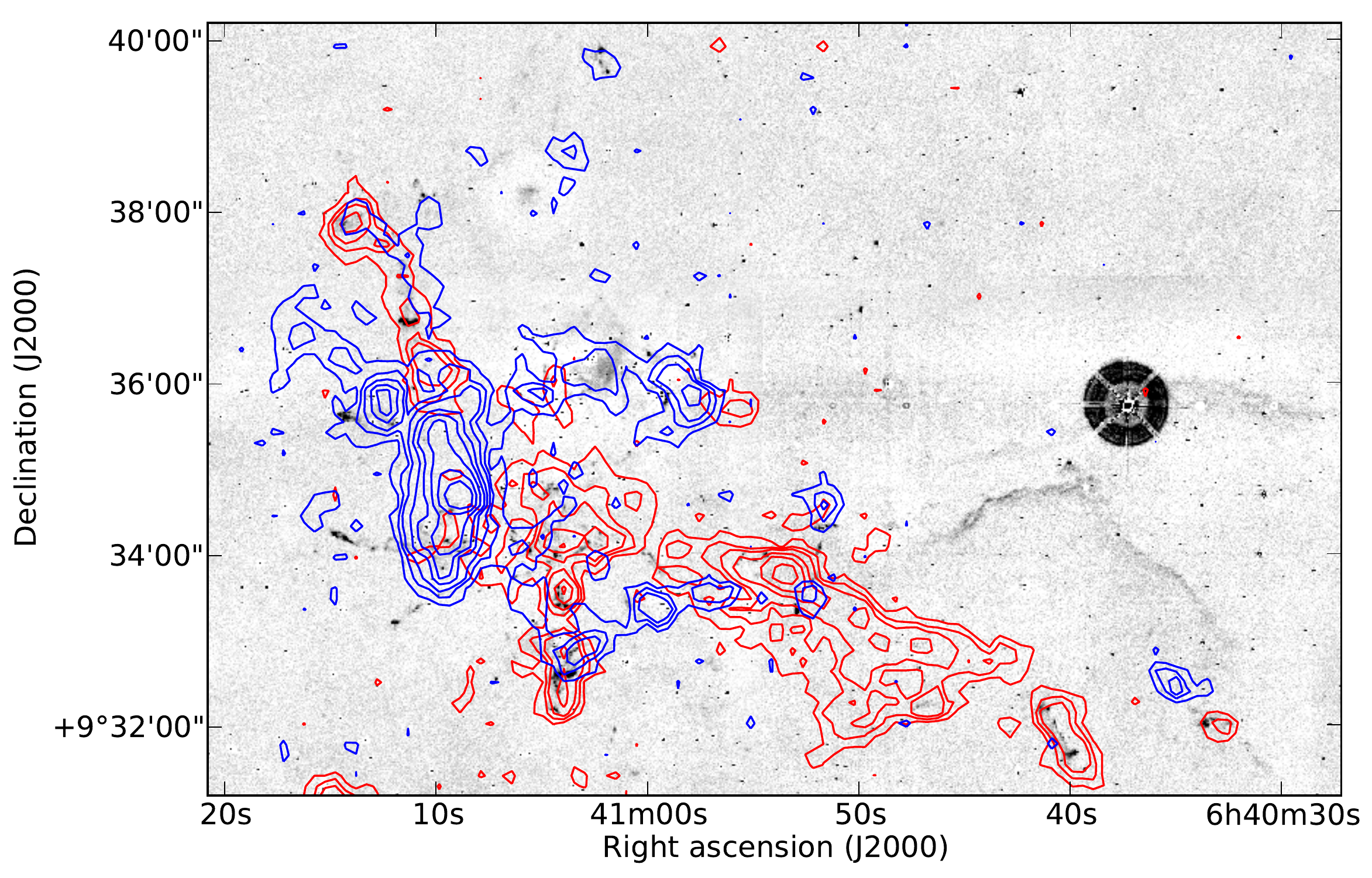}
  \caption{Red and blue shifted CO emission contoured over the H$_2$ continuum-subtracted greyscale image towards the NGC\,2264~D cluster. Contours are at 3.9, 6.5, 9.1, 14.3, 19.5, 27.3, 37.7, 48.1~K\,\kms.  Red-shifted emission extends from 11.1 to 25.5 \kms, and the blue-shifted from -18.5 to 1.0 \kms. }
              \label{fig-outflow3_main}
    \end{figure*}

Fig.~\ref{fig-outflow3_main} shows red- and blue-shifted CO emission contoured over the H$_2$ image towards NGC\,2264~D. The cluster extends further westwards than previously observed, with the discovery of new jets and molecular outflows spatially adjacent to this cluster. The red-shifted emission contoured at the south-eastern edge of the map is from the NGC\,2264~C cluster. The total gas mass is 1310~\msol\ in NGC\,2264~D \citep{peretto2006}, so the molecular outflows, entraining 5~\msol, constitute a negligible fraction ($\sim$0.4\%) of the cluster mass.  Although this is a lower limit, this result implies that the large number of outflows are not affecting the bulk of molecular material.

\citet{sung2009} classifies 44 sources as protostellar in NGC\,2264~D.  There are many overlapping CO outflows and H$_2$ jets, which we are unable to separate near the sources, making source identification difficult. \citet{frobrich2010} identifies 3 embedded YSOs as candidate massive YSOs using PAH emission. One of these, SSB\,11829 is co-incident with a $^{12}$CO bipolar molecular outflow, and MHO\,1385, a bright knot and extended west-facing bow shock, indicating the source is protostellar. A more detailed description of the individual H$_2$ objects and CO flows is given in the Appendix.

\subsection{NGC\,2264~G}

\begin{figure}
   \centering
   \includegraphics[width=8.5cm,angle=0]{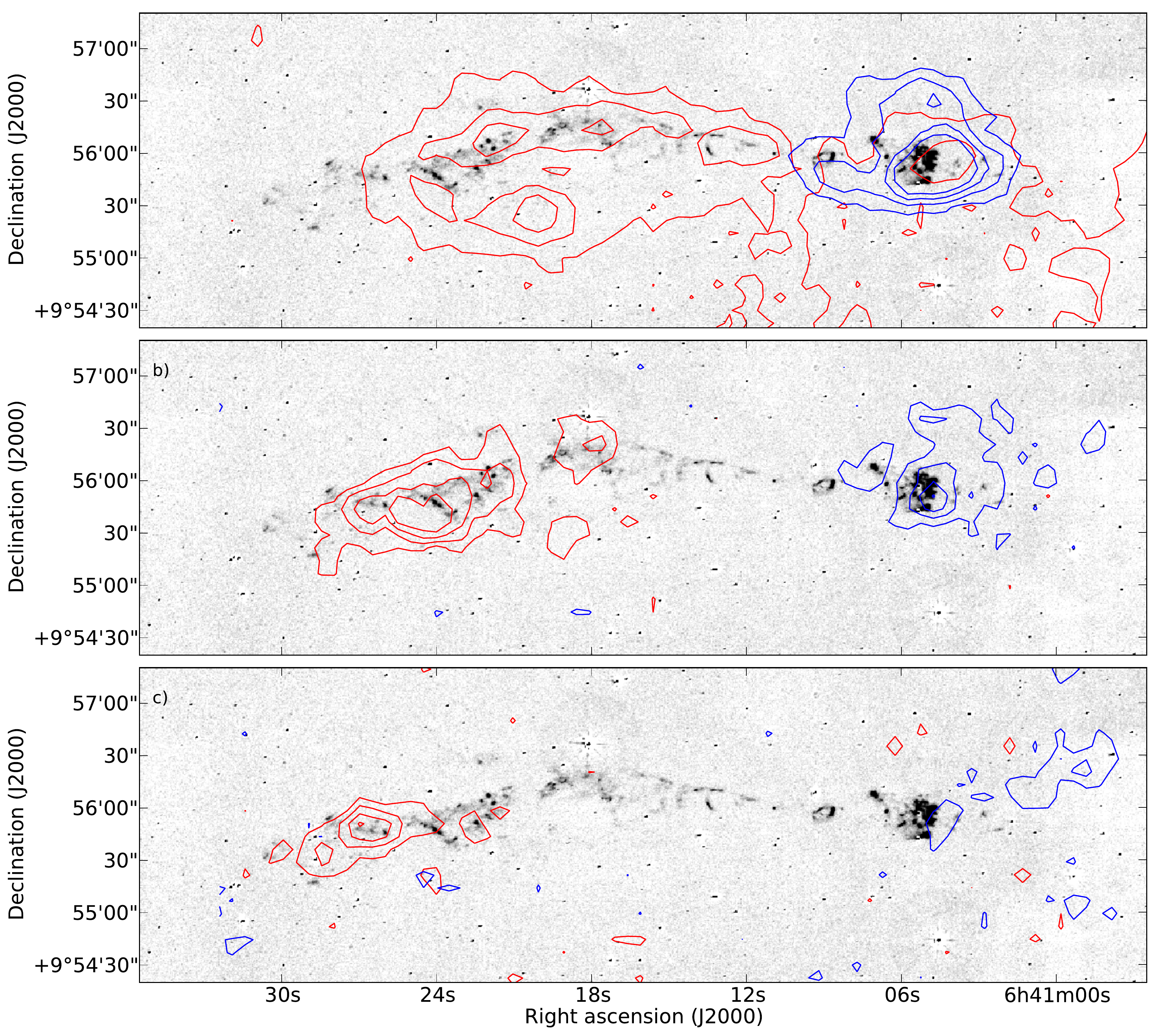}
  \caption{Red- and blue-shifted CO emission contours for three different velocity ranges towards NGC\,2264, overlaid on the H$_2$ greyscale image. The red-shifted emission is associated with MHO 1359, the blue-shifted emission with MHO 1358. (a) Low-velocity gas at 6.0--8.0 \kms\ (red) and 1.0--3.0 \kms\ (blue). CO contours start at 2.5 K\,\kms, with steps of 2.5 K\,\kms. (b) Mid-velocity gas at 16.0--26.0 \kms (red) and -17.0--\,-7.0 (blue), contours as (a).  (c) High-velocity gas at 29.0--39.0 \kms\ (red) and -30.0--\,-20.0 \kms\ (blue). CO contours start at 2.0 K\,\kms, with steps of 2.0 K\,\kms. }
              \label{fig-oc1}
    \end{figure}

The two outflow lobes of NGC\,2264~G can be seen just to the northeast
of S\,Mon in Fig.~\ref{fig-wco}. The driving source \citep{teixeira2008,gomez} has a mass, including the envelope, of 2--4~\msol. \citet{teixeira2008} report {\it Spitzer} imaging of the protostellar jet, showing three changes in direction.  Fig.~\ref{fig-oc1} shows the red and blue contours for three different velocity ranges through the outflow lobes, overlaid on the H$_2$ emission from the protostellar jet.  The CO emission from the molecular outflow follows the emission in the jet, including the changes in direction. There is a strong correlation between the structure of H$_2$ emission and high velocity CO outflow lobes in the NGC\,2264~G outflow.  The extent of the H$_2$ emission matches the size of the CO lobes, and the H$_2$ intensity peaks are close to the peaks seen in the CO maps.

The total extent of the red outflow lobe is $\sim$280$\arcsec$, or 1.1~pc (assuming 800~pc to NGC\,2264), with the very high velocity gas ($>$35~\kms) extending further than the lower velocity gas. The low velocity emission (6~\kms) traces a cavity, while the high velocity emission (40~\kms) traces a collimated flow. The blue lobe has a small high velocity extension, and the total extent of the blue outflow lobe seen is $\sim$200$\arcsec$, or 0.8~pc.  These figures are in agreement with those derived from CO 2$\rightarrow$1 observations \citep[]{lada}. The difference in the highest velocity spatial extent between the red and blue lobes we detect in CO 3$\rightarrow$2 may be merely the amount of material available for entrainment in the two directions. The red lobe in Fig.~\ref{fig-oc1}(a) suggests a wide opening angle, with the H$_2$ emission (MHO 1359) associated with the northern edge of the flow.  The clumpy structure seen very clearly in our new images of MHO 1359 could result from shocks fronts in the boundary layer between the flow and the ambient medium.   The lack of H$_2$ emission along the southern edge of the flow lobe could be simply be due to a lower ambient density, suggesting a gradient in the ambient density that increases to the north across this region.  Alternatively, \citet{teixeira2008} propose a slowly precessing jet, plus additional deflection, to explain the structure.  The broad red lobe we detect may suggest a density gradient, rather than a precessing jet. Table \ref{tab-mass} gives the calculated outflow parameters for the NGC\,2264 outflow. The total mass entrained in the outflow is 0.9~\msol\, with a momentum of 11~\msol~\kms. This is consistent with the mass  and momentum found for the lower excitation CO $2\rightarrow1$ transition \citep[1~\msol, 12~\msol~\kms, ][]{lada}. The jet is bright, and has the highest $L_{H_2}$ luminosity for an individual flow in this region. 

\subsection{S\,Mon}
S\,Mon is positioned in front of and travelling towards the NGC\,2264 cloud \citep*{tauber}, affecting the region with an intense ionizing field. The emission surrounding S\,Mon shows many filaments and arcs, although we do not detect any evidence from the presence of extended line wings, or the presence of H$_2$ knots, any protostellar outflow activity. The emission near S\,Mon is seen across a large velocity range, from 6.7~\kms\ to 13.5~\kms, and is characterized by relatively small linewidths, indicating that the complex kinematic structure is not due to protostellar outflow activity. At the highest red and blue velocities, the emission is very clumpy and fragmented. In the ring of emission (Fig.~\ref{fig-wco}), a cluster of SCUBA cores to the north-west is associated with CO clumps at blue-shifted velocities, while the SCUBA cores to the south-east are associated with CO clumps at red-shifted velocities. 

\subsection{NGC\,2264~IRAS-2}
\begin{figure}
   \centering
   \includegraphics[width=6cm,angle=0]{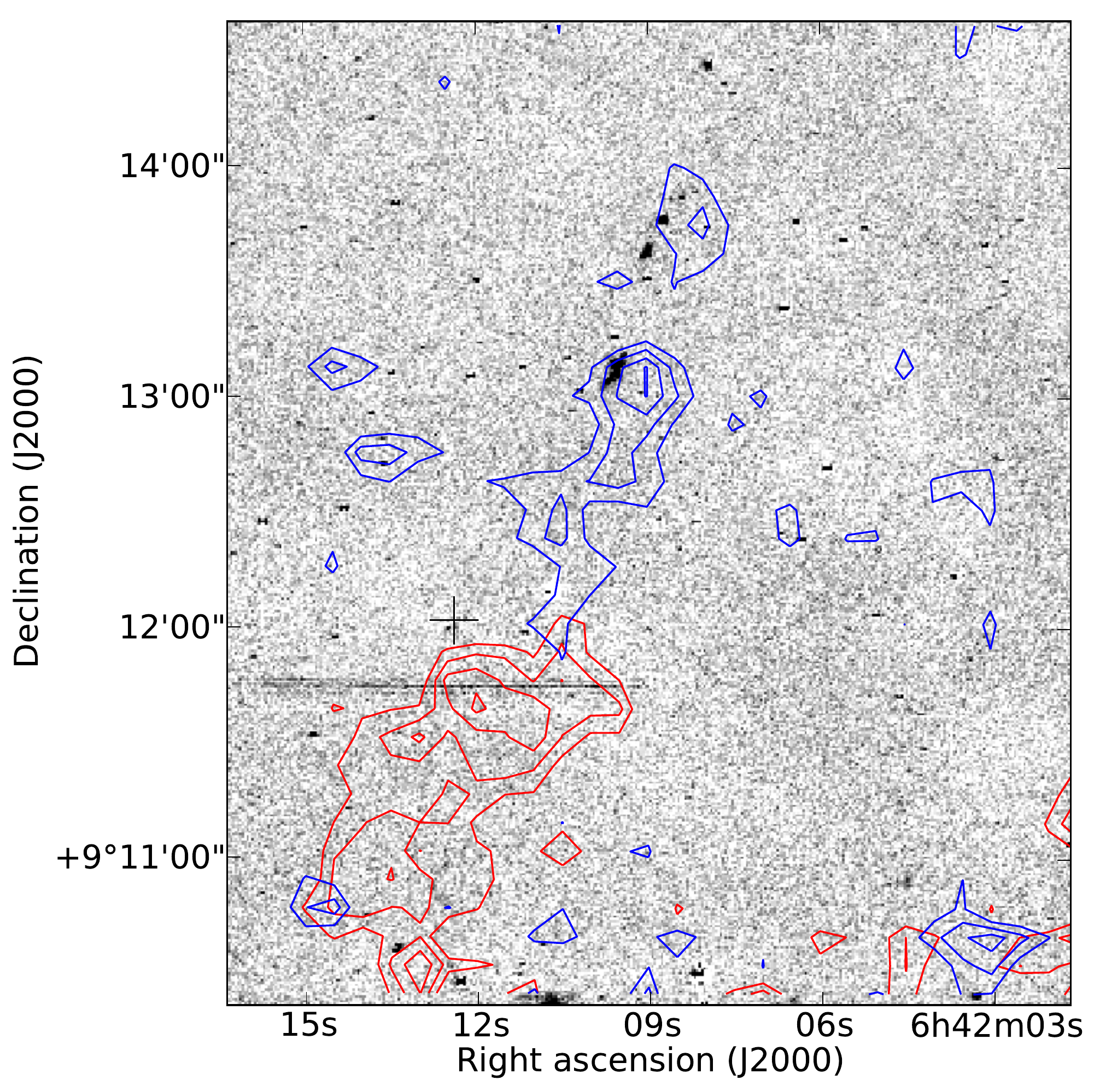}
  \caption{Red- and blue-shifted CO emission contours, overlaid on the H$_2$ greyscale image towards MHO 3109, shown as an ellipse. Red CO contours start at 3.0 K\,\kms\ with 1.0 K\,\kms\ steps, and blue contours start at 2.0 K\,\kms\ with 0.8 K\,\kms\ steps. A cross marks the position of the far-IR source IRAS2 \citep{margulis}. }
              \label{fig-new_south}
    \end{figure}

Fig.~\ref{fig-new_south} shows outflow lobes associated with a source first detected by \citet*{margulis}. This is a far-IR source with no optical counterpart and no previously detected outflow, designated as IRAS-2, with a luminosity of 6.7~\lsol. \citet*{wolfchase1995} detected an extended CS source at this position, but no outflow. Along with a weak CO molecular outflow, we also detect H$_2$ knots indicating protostellar jet activity, in MHO 3109 (Fig.~\ref{fig-new_south}). 

The outflow candidate NGC\,2264\,A \citep*{margulisb} is $\sim$400\arcsec\ east of IRAS-2. Although a red outflow lobe has previously been detected towards NGC\,2264\,A, we do not clearly detect the outflow lobe in CO~$3\rightarrow$2 emission, or detect any H$_2$ emission.

\subsection{IRAS~06396+0946}

\begin{figure}
\includegraphics[angle=0,width=4cm]{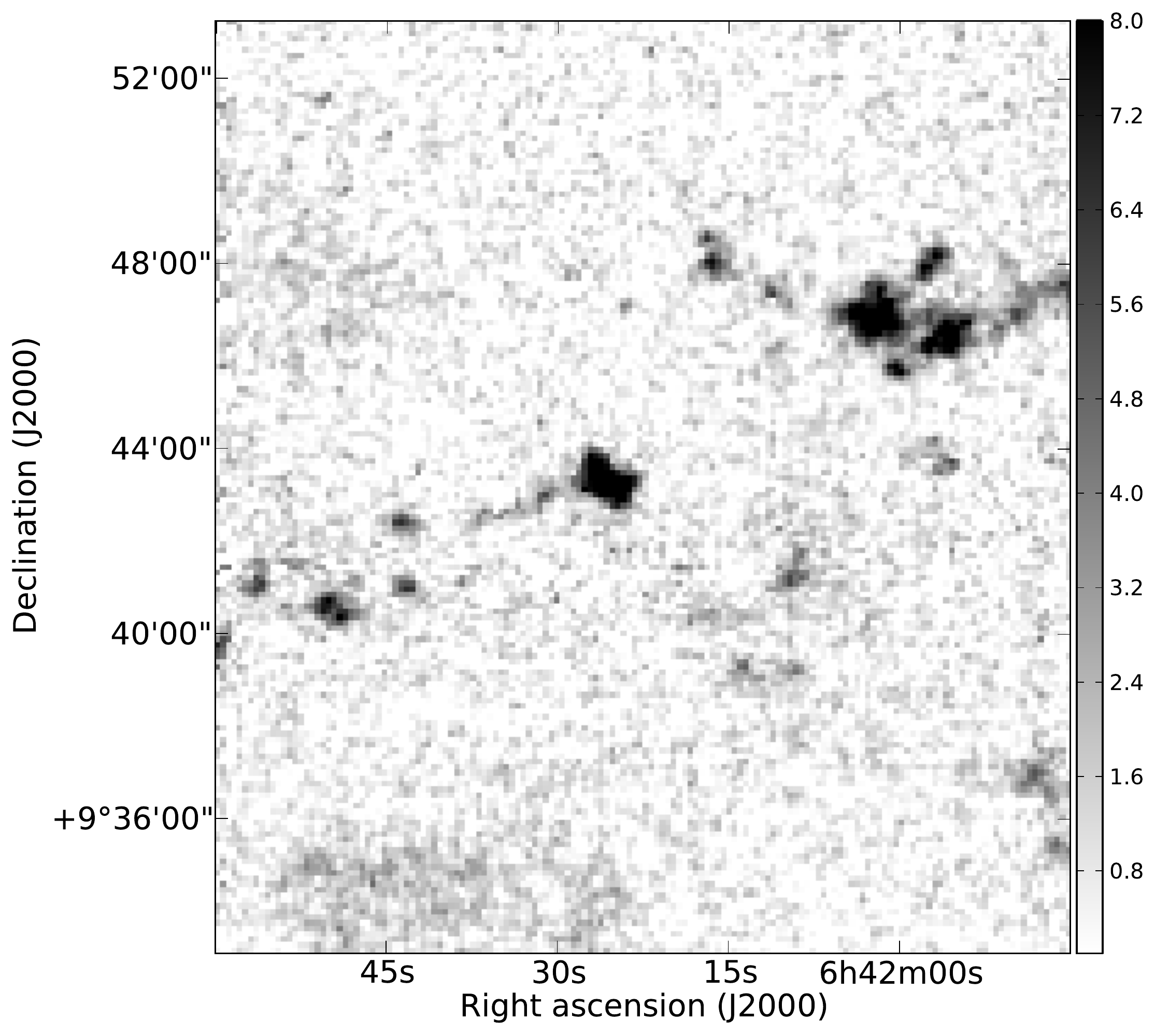}
\includegraphics[angle=0,width=4cm]{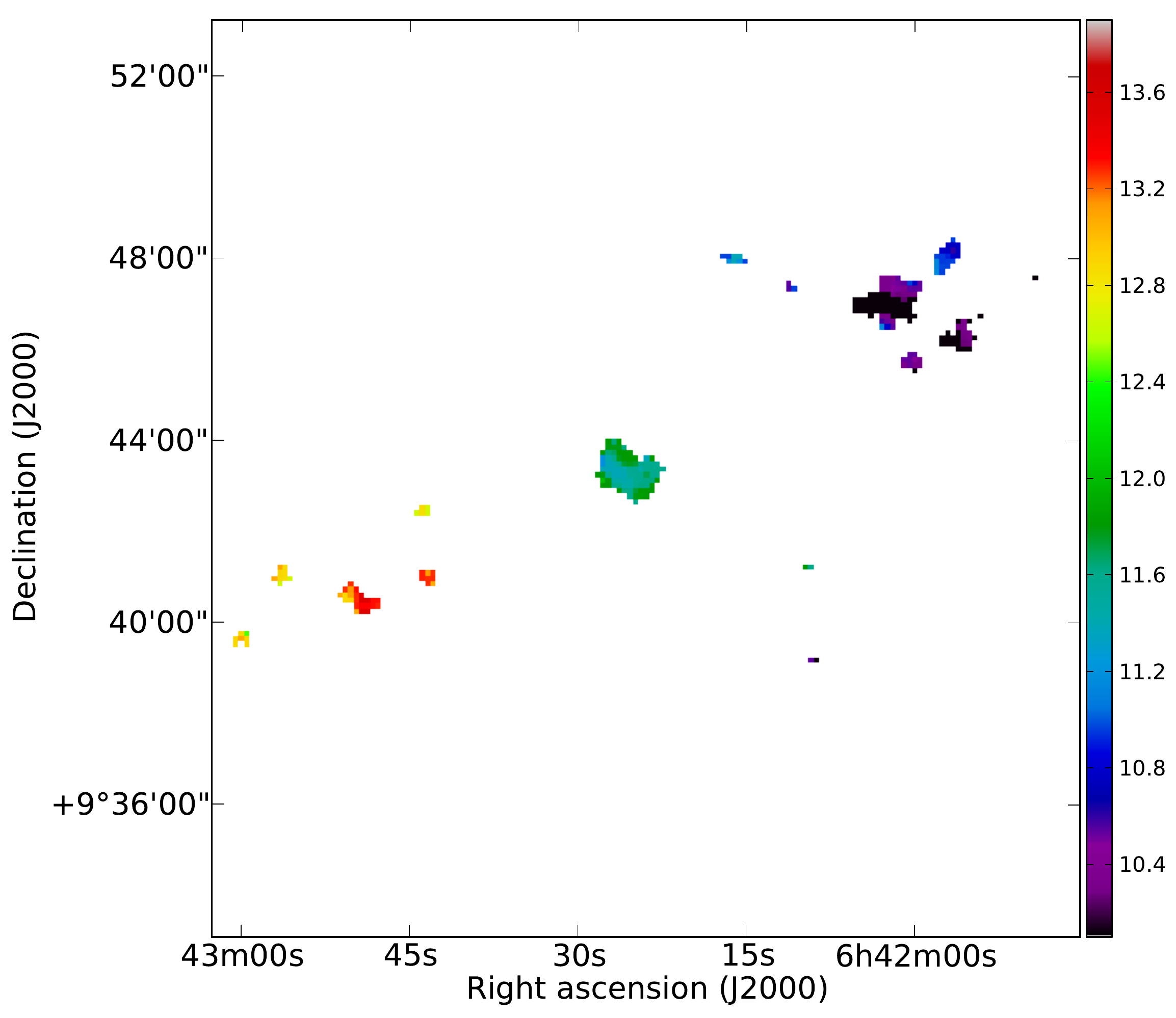}
\caption{The isolated cometary feature east of the main NGC\,2264 filament. Left: $^{12}$CO integrated intensity in greyscale. Right: The velocity field shown through the intensity-weighted velocity image. The velocity data has been thresholded at the 3$\sigma$ level for clarity. \label{fig-east}}
\end{figure}

East of the main emission region, at the position 06$^{\rm h}$42$^{\rm m}$25$^{\rm s}$.5 +09$^{\arcdeg}$43$^{\arcmin}$09$^{\arcsec}$ is a cometary-shaped feature, surrounded by several small, compact regions in $^{12}$CO,  which are blue-shifted to the north-west, and red-shifted to the south-east. The peak antenna temperature is 10.8~K, and the lines are narrow, with FWHM $\sim$1.0~\kms\ from a Gaussian fit to the data. Fig.~\ref{fig-east} shows the integrated intensity image of this region, in the velocity range 10.1~\kms\ to 13.9~\kms\, along with the intensity-weighted velocity image, which shows the velocity field. This is an intriguing structure, which could be associated with IRAS~06396+0946, although there is as yet no other published data on this region. The velocity field suggests an outflow or jet, although the morphology of the $^{12}$CO emission is more reminiscent of H$_2$ knots, such as those in the blue-shifted jet of NGC\,2264~G, rather than the more linear or elliptical structures more commonly seen in CO. There is no H$_2$ emission associated with this object. The spatial alignment and velocity structure of this region may just be coincidental, however, with the emission arising in areas at different distances which are not related. More sensitive observations would be required to determine whether these emission regions are connected.

\subsection{The NGC\,2264~D filament}

To the east of NGC\,2264~D, there are two newly detected MHOs, MHO1378 and MHO1379, shown in the Appendix (Fig.~\ref{h2g}b). The location of the MHOs and the velocity structure of the CO emission suggests that these MHOs may be associated with molecular outflows which are erupting from the dense gas in the cluster, which contains several {\it Spitzer}-identified protostars \citep{sung2009}. However, the line profiles do not have the line wings usually associated with molecular outflows. As the velocity increases from 1.0~\kms, two filaments extend eastwards from the bulk of the cluster, eventually overlapping and merging at velocities $\sim$5.0--6.5~\kms, when the overall length is $\sim$11\arcmin. At higher velocities, the filament becomes more pronounced, with additional internal structure that is very linear. The brightest regions are those furthest from the central cluster.

\section{Principal Component Analysis}
In order to undertake an analysis of the turbulent characteristics of NGC\,2264, we have carried out a principal component analysis (PCA) on the $^{12}$CO data.   
The method is relatively robust against the effects of resolution and noise. The low-order components contain features that contribute most to the variance of the data, while higher-order components contain more subtle features within the spectral shapes.

We have implemented the technique described by \citet{heyer,brunta} and \citet{brunt} who detail the methods in full.  This technique transforms the original spectroscopic data cube (RA, Dec and velocity) into a set of orthogonal functions which are described by the principal components, $l$, ordered by decreasing variance of the data projected onto each orthogonal vector.  The principal components consist of an eigenvector, tracing only the velocity structure, and an eigenimage, constructed from the projection of the data onto each of the eigenvectors. Each eigenimage therefore contains only spatial structure, mapping the size scales of differences in the line profiles traced by the eigenvectors. The principal components have characteristic length scales ($L_l$) determined from the eigenimage, and characteristic velocity scales ($\delta v_l$) determined from the eigenvectors.  The eigenvectors and eigenimages of each principal component are coupled, so the physical dynamics can be determined from the characteristic scales. The data can be reconstructed from a linear combination of only the significant principal components, which account for most of the variance, or variability, in the data.

The technique provides an objective method of extracting the most significant components of the data for analysis, the challenge then being to interpret the results in a physically meaningful way. \citet{brunta} have carried out PCA on simulated data in order to provide empirical relations between the derived PCA results and the statistics of the original data, which we utilize in this analysis to aid the physical interpretation.

\citet*{brunt09} used numerical magnetohydrodynamic models and molecular spectral line observations to show that the ratio of the characteristic length scales of the first two principal components ($L_2/L_1$) is related to the ratio of the turbulent driving scale to cloud size, ($\lambda_D/L_D$) for isotropically forced turbulence \citep[see also][]{brunt2003}. Therefore, observational measures of $L_2/L_1$ can be used to estimate $\lambda_D/L_D$. For ratios $>0.2$, there is little sensitivity to the actual driving scale, and large-scale driving best describes the turbulent driving scale \citep{brunt09}.

\subsection{Principal Components}

Following \citet{brunt2003}, we do not subtract the mean from the data, so that $l_1$ approximates the integrated intensity and mean line profile of the $^{12}$CO data. The scale of data values in the eigenvectors is not directly related to flux, but to the contribution of the principal component to each velocity channel. Large relative positive or negative values indicate that a principal component contributes significantly in that velocity channel. The eigenimages can be used as a diagnostic of how much spatial structure is associated with each eigenvector.

Table~\ref{tab-eig} lists the first 10 principal components, along with the variance of and the total cumulative variance for each component. The first five principal components contribute $>$99\% of the variance, with $l_1$ accounting for 74\% of the variation in the data.  The largest values of $L_l$ corresponds to the largest values of $\delta v_l$ indicating that the largest velocity differences are distributed on the largest scales.

\begin{table}
\begin{tabular}{rrlll}
\hline
$l$& Variance &Cumulative Variance&$L_l$&$\delta$\,$v_l$\\
&\%&\%&pc&\kms\  \\
\hline
1&  74.15 &74.15&7.5& 6.3\\
2&  20.86& 95.01&2.6 &2.5\\
3&   3.16& 98.18&1.3 &1.7\\
4&   0.73 &98.90&0.6 &1.3\\
5&   0.51& 99.41&0.6 &0.8\\
6 &  0.21& 99.63&&\\
7&   0.08& 99.71&&\\
8&   0.05& 99.76&&\\
9 &  0.03& 99.79&&\\
10&  0.02& 99.82&&\\
\hline
\end{tabular}
\caption{Contribution of the first 10 principal components to the NGC\,2264 CO data. Characteristic length and velocity scales for the 5 significant principal components are listed. \label{tab-eig}}
\end{table}

\begin{figure*}
\includegraphics[angle=0,width=12.5cm]{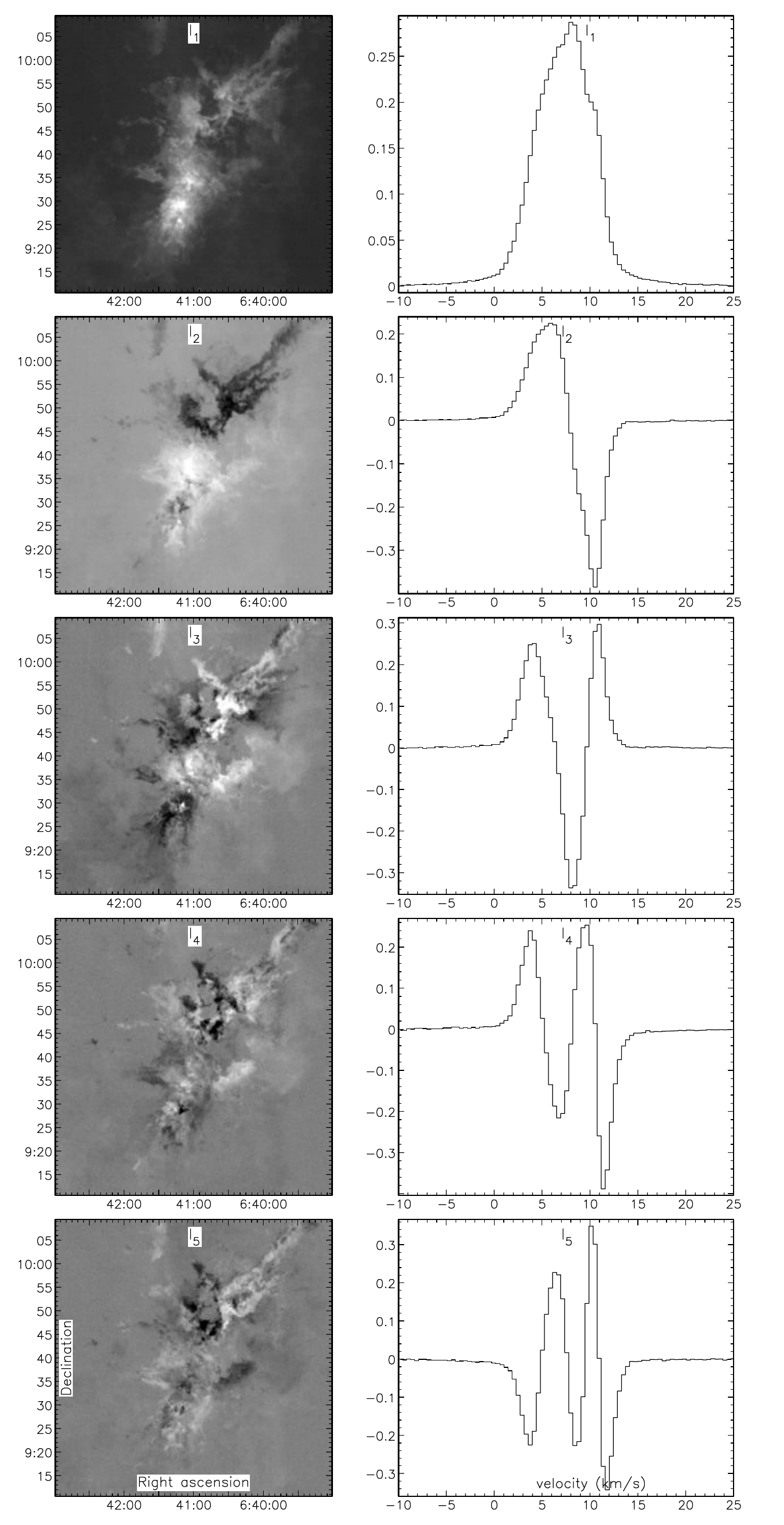}.
\caption{Eigenimages and eigenvectors of principal components 1 to 5.\label{fig-eig1}}
\end{figure*}

Fig.~\ref{fig-eig1} shows the eigenimages and eigenvectors associated with the first 5 principal components. The black and white (positive and negative) regions in the eigenimages trace velocity fluctuations whose magnitude is indicated by the width of features in the eigenvectors. The first principal component, $l_1$, as noted above, has an eigenimage which approximates the $^{12}$CO integrated intensity map. The centre velocity of the emission is 7.4~\kms, and it has a large an FWHM linewidth of 6.9~\kms, from a Gaussian fit to the eigenvector, which approximates the mean line profile of the $^{12}$CO data.  
 $l_2$ shows a positive-negative dipole pattern that is characteristic of giant molecular clouds \citep{brunt09}, when the turbulence is driven on scales comparable to the cloud size.
The lower velocity contributes mostly to spatial variations in the north, while the higher velocity contributes mostly to spatial variations in the south. This is consistent with the velocity field we determined for NGC\,2264, discussed in Sec.~\ref{obs-summ}. The positive velocity component of $l_2$ is at 5.4~\kms, with a linewidth of 4.2~\kms. This is spatially most closely associated with the emission from the star-forming clusters in the cloud, NGC\,2264~C and NGC\,2264~D. The negative velocity component of $l_2$ is at 10.4~\kms, and has a narrower linewidth of 2.6~\kms. This component is spatially most associated with emission from the region near S\,Mon, and the tip of the Cone. $l_3$, with peaks at 4.0, 8.2 and 10.9~\kms, and linewidths of 2.9, 1.7 and 1.6~\kms,  describes velocity and linewidth variations within the velocity components identified in $l_1$ and $l_2$. All of the significant principal components also show an isolated, compact region of emission to the east of the main emission region, at the position associated with IRAS~06396+0946 (discussed in Sec.~\ref{outflows}).  

Further principal components describe successively smaller spatial variations in velocity and linewidths. By $l_{20}$, there is negligible emission in the eigenimages, and from these high-order principal components, noise estimates can be made \citep{brunt}. 

\subsection{Characteristic Scales}
Using results from the PCA decomposition, we can investigate the characteristic velocity variations and the spatial scales over which they occur. The characteristic scales are calculated using the normalized autocorrelation functions (ACFs) of each principal component, $l$, following \citet{brunta}. The spatial scales are calculated using noise-subtracted, resolution-corrected ACFs. Given the limited number of spectral pixels, the characteristic velocity scale measurements are largely unaffected by instrumental noise. Instrumental noise and finite resolution effects are explained in detail by \citet{brunta}, whose procedures we have implemented. The characteristic velocity and length scales ($\delta v_l$ and $L_l$) are determined from the lag at which the normalized ACFs of the eigenvectors and eigenimages fall to $e^{-1}$, and are calculated following \citet{heyer}.

The characteristic length scales $L_l$, and the characteristic velocity scales $\delta v_l$ for the 5 significant principal components are listed in Table~\ref{tab-eig}. Both the length and velocity scales decrease with increasing $l$. These scales can be visualized in the eigenimages and eigenvectors, where we can see, respectively, the spatial scale of variation, and the width of spectral features, decreasing as $l$ increases. The assumption of a PCA analysis of molecular cloud spectral line data is that the decreasing size scales are reflecting a fundamental property of the kinematics of the gas within the cloud \citep{heyer}, from which turbulent driving scales ($\lambda_D$) can be determined. The longest flow we detect towards NGC\,2264 is that of NGC\,2264~G, at 280$\arcsec$, or 1.1~pc. Comparing this value with the $L_l$ size scales suggests that the outflows we detect are contributing to the variance in the data at characteristic scales $<L_3$, and so are contributing $<$3\% to the variance in the data. They are not, therefore, contributing at the size scales necessary to be the main component driving turbulence. This is supported by our analysis in Sec.~\ref{outflows}, where we found that $<$0.5\% of the cluster masses have been entrained within molecular outflows.

On the largest scales, the velocity variations are due to differences between components that make up the entire molecular cloud. The characteristic spatial scale of the first eigenimage, $L_1$=7.5~pc, therefore gives an estimate of the overall cloud size, $L_D$. The characteristic scale of the second eigenimage, $L_2$=2.6~pc, measures the size-scale over which the largest velocity variations occur in the data. $L_2$=2.6~pc can therefore be associated with the turbulent driving scale. 

\citet*{carroll} have examined the use of PCA in simulations of outflow driven turbulence, isotropically forced turbulence, and a combination of both. They find that the ratio of characteristic length scales $l_2/l_1$ correctly identifies the largest driving scale in all three types of simulations, but suggest the scales derived from PCA are measuring the largest scale of coherent motion, which is not necessarily the turbulent driving scale. As noted by \citet{carroll}, this distinction is important in clouds where outflows are present within an external cascade, or where there are large-scale coherent motions due to angular momentum conserving collapse. In our PCA analysis, the principal components describe a large-scale velocity field across the cloud, with contributions from high velocity outflow emission. As shown in Fig.~\ref{fig-avspec}, the velocity field of NGC\,2264 shows little evidence for large-scale coherent motions. There is no indication of a rotation axis for the region as a whole.

We follow \citet{brunt09} and use the ratio of the characteristics length scales, $l_2/l_1$, to estimate the turbulent driving scale.  From the ratio of the spatial scales of the first two eigenimages, we derive a fractional driving scale of 0.35, indicating large-scale driving. Combined with the characteristic scale $L_2$=2.6~pc associated with the turbulent driving scale, and our analysis in Sec.~\ref{outflows}, these data indicate that protostellar outflow activity in NGC\,2264 is not the dominant component of turbulence, even though it is a young and energetic star forming region. 

\subsection{Size-linewidth correlation}

\citet{larson} developed an empirical relation between the linewidth $\Delta V$ and cloud size $R$, with $\Delta V \propto R^{\alpha}$, and $\alpha$=0.5. This power-law relation is similar to that expected from studies of turbulence, where values of $\alpha$ can range from  $\alpha=1/3$ for the incompressible energy cascade to $\alpha=1/2$ for a shock-dominated turbulent fluid \citep[][ and references therein]{mckee2007}. 
The equivalent PCA velocity statistic $\alpha$ is obtained from the set of significant $\delta v_l-L_l$ pairs which are larger than the resolution limits. These form the power-law relationship $\delta v \propto L^{\alpha}$. \citet{brunt2003} compared observations to simulations, and obtained values of $\alpha$ from 0.5 to 0.8. 

\citet{brunt2003a} relate the exponent of the characteristic velocity and length scale relation, $\alpha$,  to the scaling properties of the intrinsic velocity field. \citet{roman} derive a PCA calibration for the exponent of the turbulent velocity spectrum $E(k)\propto k ^{- \beta_v}$, finding $\beta_v = 0.2\pm0.05\,+\,(2.99\pm0.09)\alpha$. Values of $\beta=5/3$ are expected for the incompressible energy cascade model \citep[][and references therein]{elmegreen}. \citet{roman} applied their results to a sample of 367 molecular clouds, finding $\langle \alpha \rangle = 0.61\pm0.2$, and $\langle \beta \rangle = 2.06\pm0.6$.

We calculate $\alpha =0.74\pm0.08$ and $\beta =2.4$ from our PCA analysis. The value of $\alpha$ is larger than that expected from a classic linewidth-size relationship \citep{larson} in giant molecular clouds, but is consistent with measurements towards the NGC\,2264~C region \citep{maury} using observations of $^{13}$CO. The size-linewidth correlation is thought to depend upon the mean surface density of the cloud \citep{heyer09}, and so $\alpha$ values will be higher in regions of high surface density. NGC\,2264 has a mean surface density $>$10 times that assumed for the classic linewidth-size relation \citep{maury}. $\alpha=0.74$ is consistent with values found in high mass star forming regions such as the Rosette molecular cloud \citep*{heyer2006}. 

The $\beta$ value of 2.4, giving $E(k)\propto k ^{-2.4}$, is larger than values expected for Kolmogorov-type turbulence, but is similar to values found using $^{12}$CO towards Perseus \citep{sun2006}, where values were higher in active star forming regions than in dark clouds. A steep energy spectrum is consistent with the energetic star formation activity of NGC\,2264. 

\section{Summary}

We have presented wide-field spectral imaging observations in $^{12}$CO 3$\rightarrow$2 and wide-field high resolution imaging observations in H$_2$ 1--0\,S(1) towards NGC\,2264. These observations, covering nearly 1 square degree, offer a detailed view of the star formation activity taking place in this clustered environment. 
We find that
\begin{enumerate}
\item Protostellar outflow activity in NGC\,2264 does not occur on large enough scales to drive the turbulence. The largest flows extend to $\sim$1~pc, while a PCA analysis suggests that turbulence is driven on scales larger than 2.6~pc.
\item Only a small fraction, $<$0.5\%, of the cluster gas mass in NGC\,2264~C and NGC\,2264~D, has been swept up to high velocities through protostellar outflow activity. The outflow activity is not having a sufficient impact on the cloud to be the dominant source of turbulence.
\item We detect 46 molecular jets and knots in H$_2$, 35 of which are new detections. Based on the new H$_2$ data, NGC\,2264~C and NGC\,2264~D extend to a larger spatial extent than previously determined. 
\item  NGC\,2264~D contains a larger mass of gas at higher velocities, and also more energy and momentum than NGC\,2264~C. 
\item Due to the presence of spatially distinct red- and blue-shifted CO emission, and H$_2$ emission in the form of jets or bow shocks, we characterize 4 cores in NGC\,2264~C as protostellar that were previously identified as prestellar.
\item Of three massive YSO candidates in the NGC\,2264~D cluster, we detect a bipolar molecular outflow and H$_2$ jet towards SSB~11829. 
\end{enumerate}

The large number of protostars, jets and outflows in the NGC\,2264 region, and their small separations, require very high resolution observations in order to isolate the flows and identify driving sources. This will be possible in future observational studies with the resolution and sensitivity offered by ALMA.

\section{Acknowledgments}
We thank the referee for very useful comments which have improved the paper. The United Kingdom Infrared Telescope is operated by the Joint Astronomy Centre on behalf of the Science and Technology Facilities Council of the U.K. The infrared data reported here were obtained as part of the UKIRT Service Programme, Programme I.D. U/SERV/1764. The James Clerk Maxwell Telescope is operated by The Joint Astronomy Centre on behalf of the Science and Technology Facilities Council of the United Kingdom, the Netherlands Organisation for Scientific Research, and the National Research Council of Canada. The CO data reported here were obtained as part of Programme I.D. M07BU09. This research used the facilities of the Canadian Astronomy Data Centre operated by the National Research Council of Canada with the support of the Canadian Space Agency. This paper made use of data products from the JCMT Science Archive (JSA) project. The JSA is a collaboration between the James Clerk Maxwell Telescope (JCMT) and the Canadian Astronomy Data Center (CADC).

\section{Appendix A}
\label{apa}

\subsection{Molecular Hydrogen emission-line Objects}
\label{app-mhos}
\begin{figure*}
   \centering
  \includegraphics[width=12cm,angle=0]{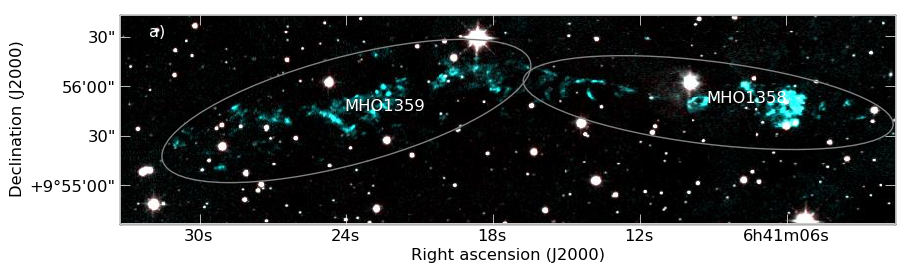}
  \includegraphics[width=8cm,angle=0]{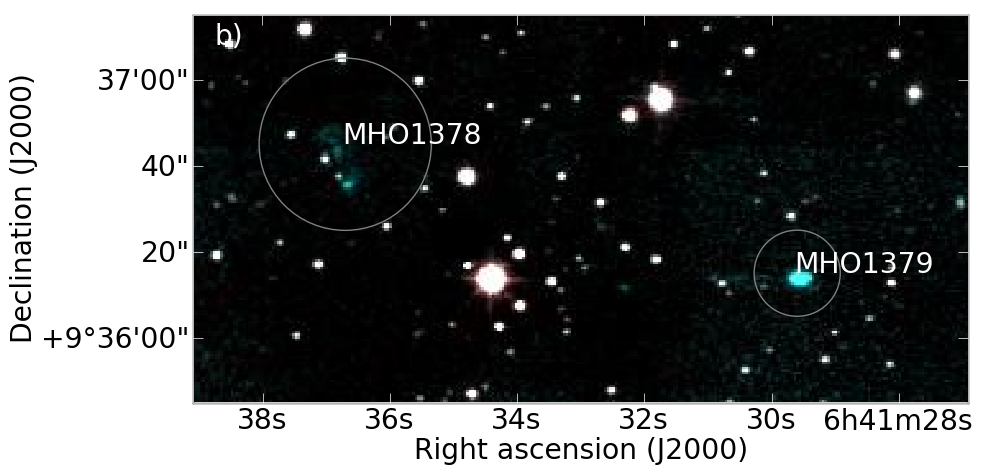}
  \includegraphics[width=6cm,angle=0]{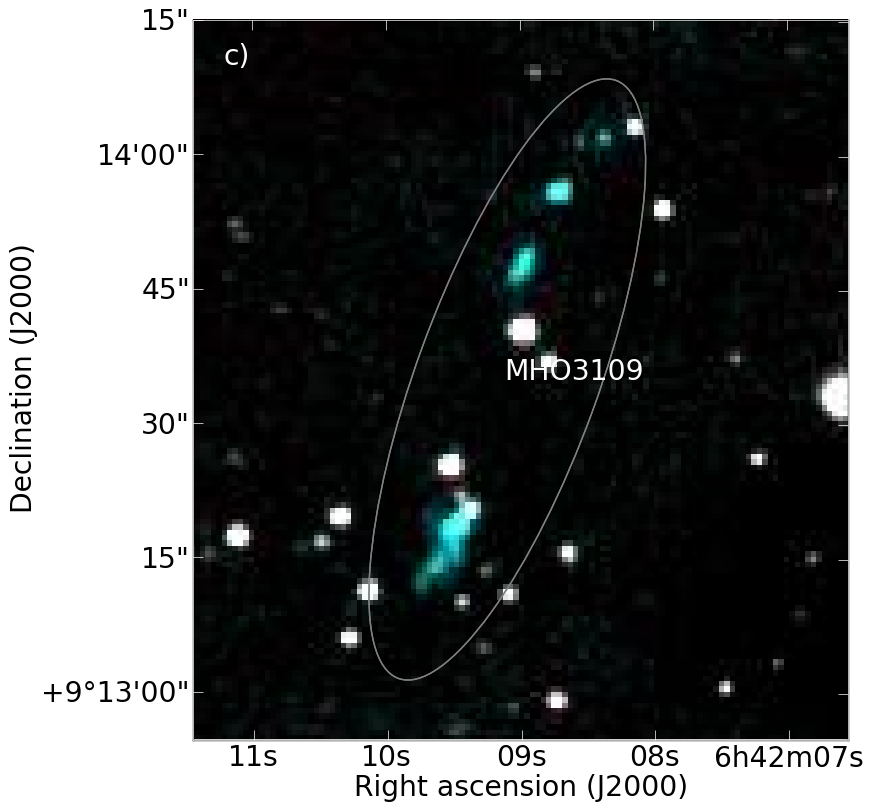}, 
 \caption{H$_2$ colour composite images (as Fig.~\ref{fig-h2}) of a) NGC\,2264~G. b) The NGC\,2264~D filament. c) NGC\,2264~IRAS2}
              \label{h2g}
    \end{figure*}

\begin{figure*}
   \centering
  \includegraphics[width=14cm,angle=0]{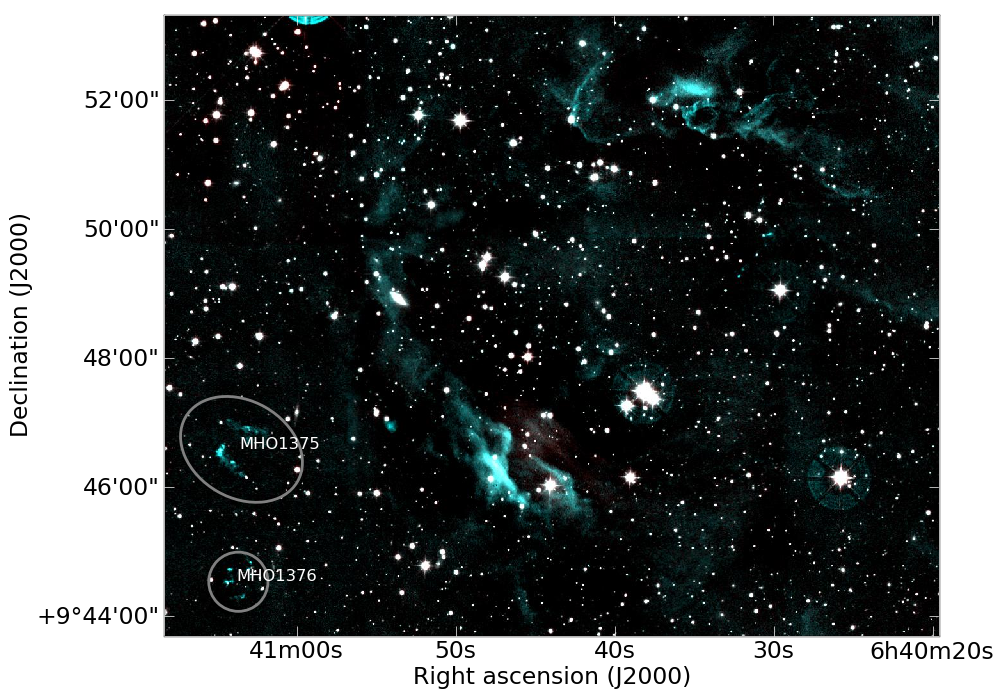}
  \caption{H$_2$ colour composite image of S\,Mon, as Fig.~\ref{fig-h2}.}
              \label{h2smon}
    \end{figure*}

Table~\ref{mhos} lists the 46 Molecular Hydrogen emission-line Objects (MHOs) identified in this work. Of these, MHO 1375-1399
and MHO 3100-3109 are new discoveries. Fig. \ref{h2g}--\ref{h2smon} show the H$_2$ emission maps of the MHOs listed in Table~\ref{mhos}. H$_2$ emission maps of NGC\,2264~C and NGC\,2264~D are shown in the next section to aid visualization when discussing the clusters.

\begin{table*}
    \caption[]{Molecular Hydrogen emission line Objects (MHOs) in NGC\,2264$^*$. }
     \label{mhos}
     $$ 
\begin{tabular}{l c c c l}
\hline
\noalign{\smallskip}
Object$^{\mathrm{a}}$ & RA         & Dec       & Assoc.$^{\mathrm{b}}$ & Note    \\
                      & (2000.0)  
& (2000.0)  & HH Object             &          \\

  \noalign{\smallskip}
  \hline
  \noalign{\smallskip}

MHO~1349   &  06 41 10.5   & +09 29 47 &       &  Emission associated with NGC\,2264~C  \\   
MHO~1350   &  06 41 09.9   & +09 28 14 &       &  Knotty jet extending to south of NGC\,2264~C/eclipsing TTS KH 15D \\   
MHO~1351   &  06 41 12.2   & +09 29 50 &       &  Arc in NGC\,2264~C region; possibly associated with MHO~1350  \\   
MHO~1352   &  06 41 15.3   & +09 29 44 &       &  Collimated jet ~1-2 arcmin ESE of NGC\,2264~C  \\   
MHO~1353   &  06 41 17.9   & +09 30 08 &       &  Chain of knots in the NGC\,2264~C region  \\   
MHO~1354   &  06 41 15.3   & +09 28 34 &       &  Compact knot in the NGC\,2264~C region  \\   
MHO~1355   &  06 41 17.7   & +09 27 59 &       &  Small knot in the NGC\,2264~C region  \\   
MHO~1356   &  06 41 12.7   & +09 28 03 &       &  Compact feature in the NGC\,2264~C region  \\   
MHO~1357   &  06 41 16.8   & +09 27 10 &       &  Small, faint knot in the NGC\,2264~C region  \\   
MHO~1358   &  06 41 05.0   & +09 55 50 &       &  Complex group of knots associated with bipolar outflow NGC\,2264~G   \\   
MHO~1359   &  06 41 23.0   & +09 55 45 &       &  Faint arcs and filaments associated with bipolar outflow NGC\,2264~G  \\ 
MHO~1375   &  06 41 04.2   & +09 46 35 & HH 125& Knots and extended bows $\sim$7\arcmin\ south of S\,Mon in NGC\,2264 \\
MHO~1376   &  06 41 03.7   & +09 44 32 & HH 225& Knots and faint emission $\sim$7\arcmin\ south of MHO~1375 (unrelated) \\
MHO~1377   &  06 41 02.2   & +09 39 51 & HH 226& Two compact knots to the north of a cluster of outflows near NGC\,2264~D  \\
MHO~1378   &  06 41 36.5   & +09 36 31 &       & Compact knot to the west of the NGC\,2264~D cluster   \\
MHO~1379   &  06 41 29.3   & +09 36 08 &       & Faint feature to the west of the NGC\,2264~D cluster    \\
MHO~1380   &  06 41 14.3   & +09 37 50 &       & Faint knots and emission in NE corner of the  NGC\,2264~D cluster   \\
MHO~1381   &  06 41 10.6   & +09 38 10 &       & Compact knot plus diffuse emission in NE corner of the  NGC\,2264~D cluster \\
MHO~1382   &  06 41 11.3   & +09 36 42 &       & Bright, knotty arc plus diffuse emission in the  NGC\,2264~D cluster   \\
MHO~1383   &  06 41 04.8   & +09 36 07 &       & Faint emission in the NGC\,2264~D cluster   \\
MHO~1384   &  06 41 02.7   & +09 35 55 &       & Faint emission in the NGC\,2264~D cluster    \\
MHO~1385   &  06 41 00.0   & +09 36 19 &       & Bright knot and extended west-facing bow shock in    \\
MHO~1386   &  06 41 14.3   & +09 35 35 &       & Bright, knot bows moving north-eastward from the NGC\,2264~D region  \\
MHO~1387   &  06 41 14.2   & +09 34 10 &       & Arcs and filaments in flow moving eastward from the NGC\,2264~D region   \\
MHO~1388   &  06 41 05.6   & +09 34 02 &       & Compact knots and fingers off emission in the NGC\,2264~D cluster   \\
MHO~1389   &  06 41 04.8   & +09 34 44 &       & Knots and emission in  the NGC\,2264~D region   \\
MHO~1390   &  06 41 06.2   & +09 33 44 &       & Extended, collimated outflow in the NGC\,2264~D cluster    \\
MHO~1391   &  06 41 04.2   & +09 33 32 &       & Bright, extended outflow that crosses MHO 1390 in the NGC\,2264~D region \\
MHO~1392   &  06 40 59.4   & +09 33 50 &       & Faint fingers of emission extending to the south-west in NGC\,2264~D   \\
MHO~1393   &  06 40 54.9   & +09 33 55 &       & Group of knots and arcs in the NGC\,2264~D region   \\
MHO~1394   &  06 40 51.9   & +09 34 18 &       & Small group of knots; possibly associated with the MHO 1395 jet \\
MHO~1395   &  06 40 52.9   & +09 33 19 &       & Bright bow shock and knotty jet to the west of the  NGC\,2264~D cluster \\
MHO~1396   &  06 41 01.5   & +09 31 57 & HH 581& Very faint outflow in the NGC\,2264~D region   \\
MHO~1397   &  06 40 48.2   & +09 32 58 &       & Knot and faint emission a few arcmins to west of the NGC\,2264~D cluster   \\
MHO~1398   &  06 40 47.2   & +09 32 13 &       & Knot and faint emission a few arcmins to west of the NGC\,2264~D region    \\
MHO~1399   &  06 40 39.8   & +09 31 41 &       & Arcs plus diffuse emission in feature to west of the NGC\,2264~D region      \\
MHO~3100   &  06 40 33.6   & +09 32 00 &       & Small group of knots plus faint emission  $\sim$1.5\arcmin\ west of MHO 1399 \\
MHO~3101   &  06 40 52.3   & +09 29 43 &       & Compact feature to south of the NGC\,2264~D cluster   \\
MHO~3102   &  06 41 01.4   & +09 30 56 &       & Compact feature to south-west of the NGC\,2264~D cluster    \\
MHO~3103   &  06 41 10.2   & +09 30 27 &       & Curving chain of knots/outflow  $\sim$1\arcmin\ north of NGC\,2264~C  \\
MHO~3104   &  06 41 19.9   & +09 29 41 &       & Knot plus faint fingers of emission $\sim$2.5\arcmin\ east of NGC\,2264~C \\
MHO~3105   &  06 41 28.8   & +09 28 24 &       & Diffuse arcs of emission  $\sim$4.5\arcmin\ ESE of NGC\,2264~C   \\
MHO~3106   &  06 41 07.4   & +09 30 04 &       & Two compact knots $\sim$1\arcmin\ north-east of NGC\,2264~C   \\
MHO~3107   &  06 41 06.8   & +09 26 54 &       & Knot plus faint emission $\sim$3\arcmin\ south of NGC\,2264~C    \\
MHO~3108   &  06 41 09.6   & +09 25 45 &       & Bright knot and arc of emission $\sim$4\arcmin\ south of NGC\,2264~C \\
MHO~3109   &  06 42 09.5   & +09 13 09 &       & Chain of half-a-dozen knots in a flow in the south of NGC\,2264   \\
  \noalign{\smallskip}
  \hline
  \end{tabular}
     $$ 
\begin{list}{}{}
  \item[$^{\mathrm{a}}$]For discovery images of MHO~1349-1357 see
  Wang et al. (2002); MHO~1358/1359 was first imaged in H$_2$ by
  Davis \& Eisl\"offel (1995).  MHO~1375-1399 and MHO~3100-3109 are
  new discoveries.
  \item[$^{\mathrm{b}}$]Associated HH object, if any (see Reipurth et al.
  2004, for details).
  \item[$^*$]For a complete list and individual images of MHOs, see:
  http://www.jach.hawaii.edu/UKIRT/MHCat/

\end{list}

\end{table*}


\subsection{The protostellar content of NGC\,2264~C}
\label{app-irs1}

NGC\,2264~C is associated with IRS1, or Allen's source, an early B star discovered by \citet{allen}, who estimated extinction towards the source as A$_v\sim$26.  The group of prestellar cores and protostars to the south are known as the cluster NGC\,2264~C \citep{margulisb,peretto2006}, where star formation has probably been triggered by activity from IRS1 \citep{thompson,williamsgarland}. The cores are forming several intermediate/high mass stars \citep{peretto2006,wardthompson}, with core masses of 1--50~\msol\ measured from submillimetre observations. Previous studies with a variety of molecular lines have pointed to the complex velocity field in this cluster, with IRS1 associated with velocities between 8--10 \kms, surrounded by several sub-clouds with velocities of 5.5--7 \kms\ \citep{schreyer}. 
The emission also extends to the west, with no obvious driving source, or any H$_2$ counterparts. IRS1 itself drives a tightly peaked, overlapping molecular outflow. There are several submm and Class I cores clustered in the region of the IRS1 source that could be driving these energetic flows.  

\begin{figure*}
   \centering
   \includegraphics[width=15cm,angle=0]{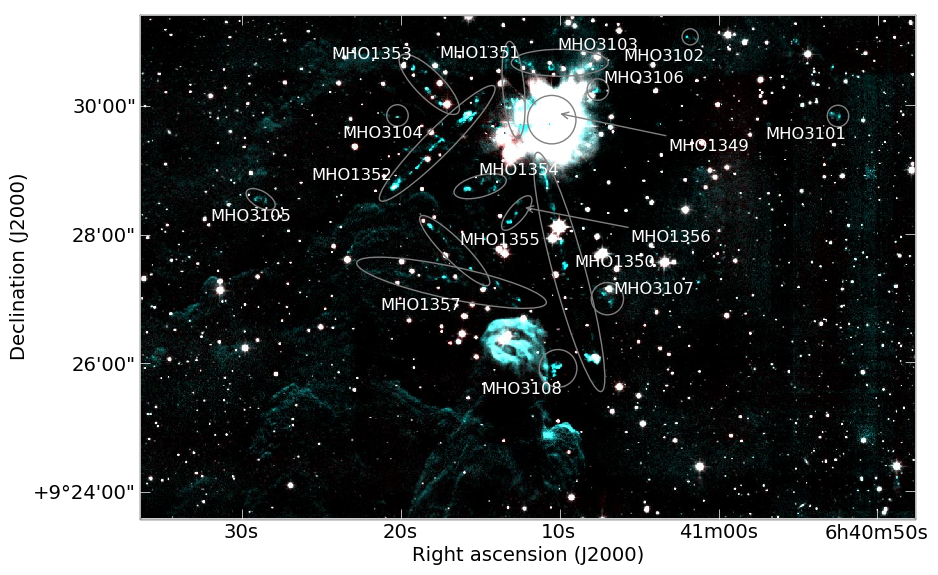}
  \includegraphics[width=15cm,angle=0]{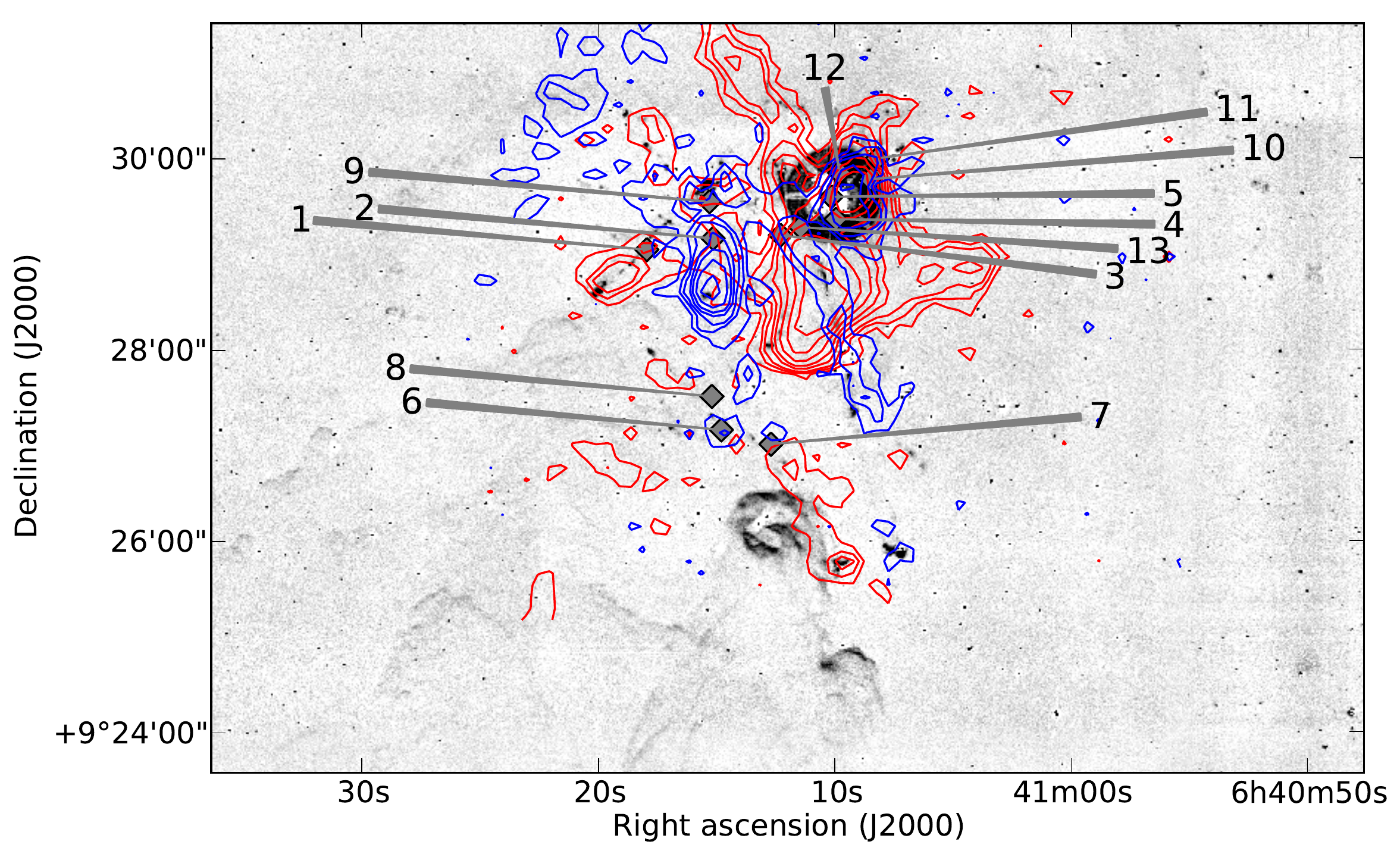}
  \caption{Top: H$_2$ colour composite image of NGC\,2264~C, as Fig.~\ref{fig-h2}. Bottom: Red and blue shifted CO emission contoured over the H$_2$ continuum-subtracted greyscale image. Contours are at 3.9, 6.5, 9.1, 14.3, 19.5, 27.3, 37.7, 48.1~K\,\kms. Red-shifted emission extends from 12.2 to 30.9 \kms, and the blue-shifted from -17.8 to 2.1 \kms. The saturated source seen in the H$_2$ emission is Allen's source \citep{allen}. The sources identified by \citet{peretto2006} are marked and labeled with the CMM number. }
              \label{fig-outflow2}
    \end{figure*}

\citet{peretto2006} identifies 4 prestellar cores, and 8 protostellar cores towards the clump near NGC\,2264~C, from observations of the 1.2~mm dust continuum, and the optically thin tracer N$_2$H$^+$. Fig.~\ref{fig-outflow2} shows the CO molecular outflows and H$_2$ jets, marked with the positions of the pre- and protostellar cores, adopting the labelling scheme of \citet{peretto2006,peretto2007}, and labelling the cores CMM1--CMM13. Of these, we detect possible outflows towards all of the prestellar cores, based on the presence of spatially distinct red- and blue-shifted emission, and H$_2$ emission in the form of jets or bow-shocks. The presence of H$_2$ jets indicates that these cores are young Class 0 sources.  The cores CMM6, CMM7 and CMM8 are located to the south of NGC\,2264~C, and \citet{peretto2006} indicates that they are not associated with any MSX or 2MASS sources, or any H$_2$ jets.  In Fig.~\ref{fig-outflow2}, CMM7 is associated with a relatively bright {\it Spitzer} 24~$\umu$m source, while CMM6 is spatially coincident with a faint 24~$\umu$m source, and all three are associated with red- and blue-shifted elongated emission which could be arising in protostellar outflows. In addition, CMM9, the northern-most source to the east of NGC\,2264~C also appears to have a molecular outflow, and be associated with a weak 24~$\umu$m source. We propose that these are all, in fact, protostellar sources, induced by the activity of NGC\,2264~C. 
Table.~\ref{tab-ngc2264-c} summarizes the CO outflows and MHOs we have been able to associate with a driving source, and the lengths of the flows, measured from either the H$_2$ emission, or from CO red- or blue-shifted emission. 

CMM1 drives a double-sided H$_2$ jet, MHO1352, and has a clearly defined red outflow lobe. The blue outflow lobe is driving towards the centre of the cluster, directly towards CMM9, and so can only be seen as a spatial extension in the blue shifted emission, aligned along the H$_2$ jet direction. 

CMM2 is at the base of a chain of H$_2$ knots and fingers, MHO1304, that is also associated with weak blue-shifted emission. We are not able to associate any red-shifted emission from this source due to the strong outflow lobes at the centre of the cluster. A {\it Spitzer} 24~$\umu$m identified protostar, 14214 \citep{sung2009} is 8\arcsec\ north of CMM2.

MHO1356 is associated with a large red-shifted outflow lobe. The jet, shows a curved chain of features, extends southwards from CMM3, and so we tentatively assign this as the driving source.  The lobes can't be traced directly back to the source due to the overlapping flows in the central regions.

CMM7, previously identified as prestellar, is at the base of a bipolar molecular outflow and associated bow shocks,  MHO3108 and MHO1354. CMM7 is also associated with a {\it Spitzer} 24~$\umu$m identified protostar, 13822 \citep{sung2009}.
CMM6 is co-incident with a chain of H$_2$ knots ending in a bow shock in MHO1355, associated with red-shifted emission. There is also a small extension of H$_2$ knots seen associated with blue-shifted emission on the other side of the source. If this jet extends any further, the emission is not distinguishable from the emission extending from the Cone Nebula. CMM8 is located between two small red- and blue-shifted emission regions which it may be driving, although there is also an H$_2$ emission knot perpendicular to the line of the outflow lobes, at a similar distance to the source. There is a {\it Spitzer}-identified protostar at this location, 14314. MHO1357 is spatially co-incident with the two Class I {\it Spitzer} identified sources, 14314 and 14415 \citep{sung2009} separated by just 11$\arcsec$. The sources and outflows in this region are closely clustered, and driving jets and outflows in different directions. This small cluster of cores (CMM6, CMM7 and CMM8) is surrounded by compact H$_2$ emission, and it is difficult to definitively assign all of the emission to one source or another. The amount of H$_2$ emission and red- and blue-shifted CO emission does suggest that these cores are protostellar, rather than prestellar as previously catalogued. 

CMM9 has been identified as prestellar by \citep{peretto2006}, but we see evidence of a chain of H$_2$ knots, MHO1353, and weak red-shifted emission associated with this source, perpendicular to the blue-shifted outflow lobe from CMM1. Any blue-shifted emission that may be associated with this source is entangled with the blue outflow lobe from CMM1. There is a {\it Spitzer} 24~$\umu$m identified protostar, 14214 \citep{sung2009} 11\arcsec\ from CMM9, which is spatially coincident with the terminal bow shock from the blue0shifted lobe of CMM1.

CMM10 is close to a {\it Spitzer} 24~$\umu$m Class I source \citep[13288][]{sung2009}, very close to IRS1, which is saturating the H$_2$ image. The {\it Spitzer} peak is spatially most closely aligned with the red outflow lobe, and is close, although not coincident with CMM10. There is also two compact  H$_2$ knots, MHO1306, which align with this source and red-shifted emission.  Therefore, we tentatively identify CMM10 as the driving source of this flow. Any blue-shifted emission is co-incident with the energetic flows at the centre of the cluster, and not easily distinguishable.

CMM13 has a spectacular bipolar jet and molecular outflow, and has a particularly clear jet and multiple bow-shocks, MHO3107, associated with the blue-shifted lobe. The red-shifted lobe is coincident with MHO1351.

The remaining CMM sources are all tightly clustered near Allen's Source, which drives an energetic bipolar flow, and is saturated in the H$_2$ emission, making it difficult to establish individual flows.

{\footnotesize
\begin{table}
\begin{tabular}{@{}l@{}rrrrrr}
\hline
Source&\multicolumn{2}{c}{H$_2$}&\multicolumn{2}{c}{CO lobes}&\multicolumn{2}{c}{length ($\arcsec$)}\\
&Red&Blue &Red&Blue&Red&Blue\\
\hline
CMM1&\multicolumn{2}{c}{MHO1352}&Y&Y&45&58\\
CMM2&&MHO1304&?&Y&&92\\
CMM3&MHO1356&&Y&?&75&\\
CMM6&MHO1355&&Y&Y&67&14\\
CMM7&MHO3108&MHO1354&Y&Y&90&100\\
CMM8&?&?&Y&Y&31&20\\
CMM9&MHO1353&&Y&?&77&\\
CMM10/13288$^a$&MHO3106&&Y&?&\\
CMM13&MHO1351&MHO1350&Y&Y&126&124\\
13350$^a$&\multicolumn{2}{c}{MHO1303}&Y&N&41&46\\
14314$^a$/14415$^a$&\multicolumn{2}{c}{MHO1357}&?&?&&\\
\hline
\multicolumn{6}{l}{$^a$ Identification from \citet{sung2009}}\\
\end{tabular}
\caption{Driving source identifications for NGC\,2264~C flows.}
\label{tab-ngc2264-c} 
\end{table}
}

\subsection{The protostellar content of NGC\,2264~D}
\label{app-irs2}

NGC\,2264~D is less luminous than NGC\,2264~C, forming several intermediate/high mass stars \citep{peretto2006} with core masses of 1.9--17.3~\msol, estimated from submillimetre observations. The submillimetre sources are less clustered than in NGC\,2264~C, but high resolution observations \citep{sung2009,teixeira2006} show many of the submillimetre sources are co-incident with multiple tightly clustered protostars.    \citet{frobrich2010} detects PAH emission surrounding 3 embedded YSOs in NGC\,2264~D, making them candidate massive YSOs. We investigate the protostellar status of these objects using our CO and H$_2$ data. All of the candidates, SSB\,1070, SSB\,11829 and SSB\,12820 are associated with compact K-band continuum sources. SSB\,11829 is co-incident with a $^{12}$CO bipolar molecular outflow, and MHO\,1385, a bright knot and extended west-facing bow shock, indicating the source is protostellar.  A second, SSB\,10710, has a very faint H$_2$ emission knot to the west, although this is not a sufficiently significant detection to be identified as an MHO, and we do not detect any clear outflow signatures in $^{12}$CO that we can identify with this source. 

\begin{figure*}
   \centering
     \includegraphics[width=15cm,angle=0]{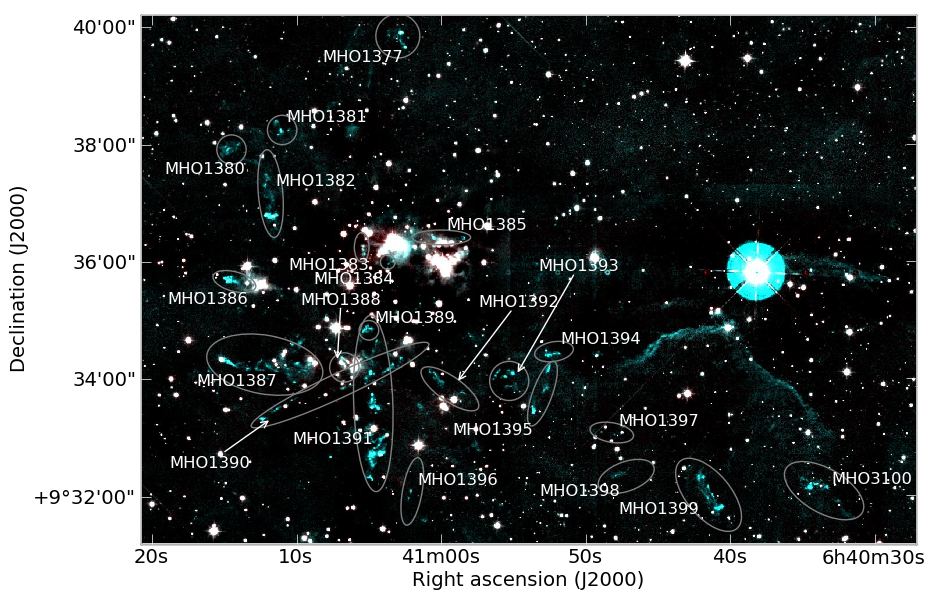}
  \includegraphics[width=15cm,angle=0]{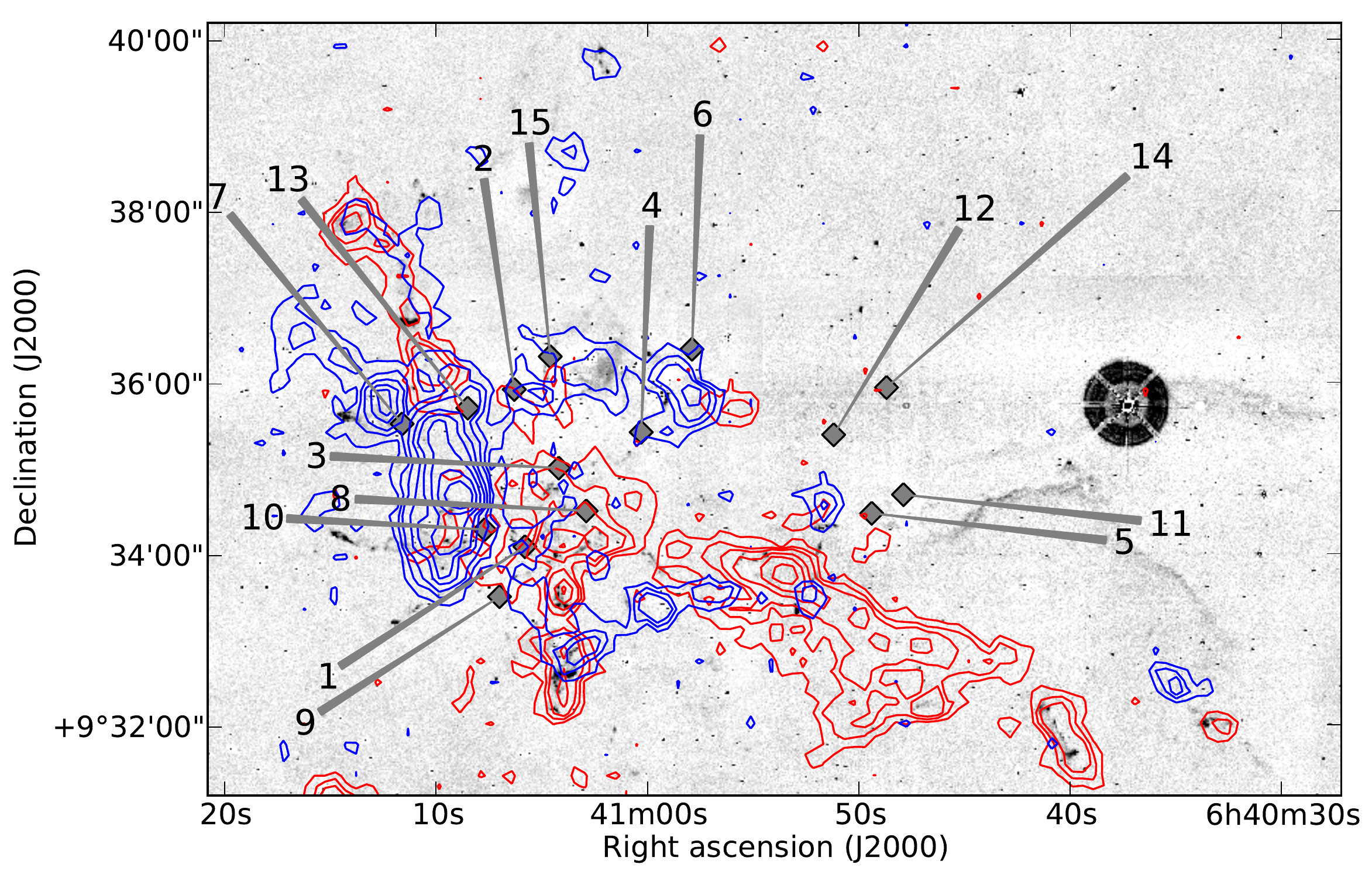}
  \caption{Top: H$_2$ colour composite  of NGC\,2264~D, as Fig.~\ref{fig-h2}.
Bottom: Red and blue shifted CO emission contoured over the H$_2$ continuum-subtracted greyscale image. Contours are at 3.9, 6.5, 9.1, 14.3, 19.5, 27.3, 37.7, 48.1~K\,\kms.  Red-shifted emission extends from 11.1 to 25.5 \kms, and the blue-shifted from -18.5 to 1.0~\kms. The sources identified by \citet{peretto2006} are marked and labeled with the DMM number.  }
              \label{fig-outflow3}
    \end{figure*}

Eleven prestellar and four protostellar cores have been identified by  \citet{peretto2006}, and we have adopted their labelling scheme of, labelling the cores DMM1--DMM15. Fig.~\ref{fig-outflow3} shows the red- and blue shifted contours overlaid on the H$_2$ emission, as Fig.~\ref{fig-outflow3_main}, with the DMM sources also labelled. Six of the prestellar cores are co-incident with {\it Spitzer} 24~$\umu$m identified protostars \citep{sung2009}, and two of these show clear outflow structure in CO emission, DMM2 and DMM13. Both of these are part of a ridge containing a previously identified protostellar source, DMM7. The final source in this ridge, DMM15 is also associated with a 24~$\umu$m source, but only has weak evidence of outflow activity. The NGC\,2264~D cluster is more tightly clustered than NGC\,2264~C, and many of the DMM sources are co-located with multiple {\it Spitzer} identified protostars, making it difficult to isolate individual flows. Where this has been possible, we list in Table.~\ref{tab-ngc2264-d} the CO outflows and MHOs we have been able to associate with a driving source, and the lengths of the flows, measured from either the H$_2$ emission, or from CO red- or blue-shifted emission. We describe the flows in more detail below.

The area surrounding MHO1385 contains two prestellar cores, DMM4 and DMM6, and 4 {\it Spitzer}-identified protostars \citep{sung2009}, one of which is co-incident with an H$_2$ knot. MHO1385 is associated with mainly blue-shifted emission, with a more compact, slightly overlapping red-shifted CO flow to the west. Although there are multiple sources in this region, the bipolar flow is not obviously associated with any of the nearby sources or H$_2$ emission.  Midway between DMM4 and DMM6 lies a candidate massive YSO, SSB\,11829 \citep{smith2010}, and it is this source that is spatially aligned with the molecular outflow and MHO\,1385. 50$\arcsec$ to the south-east of MHO1385 are two more H$_2$ objects, MHO1383 and MHO1384, two prestellar cores, DMM2 and DMM15, and 5 {\it Spitzer}-identified protostars, all within a radius of 30$\arcsec$. There is red- and blue-shifted emission associated with this region, but it is not possible to disentangle individual flows at this resolution. DMM7 is located at the base of MHO1386, which is clearly associated with blue-shifted emission, although we can't isolate any red-shifted counterpart to this flow.

MHO1380 is coincident with a compact peak of red-shifted CO emission. MHO1381 is associated with weak blue-shifted emission, and could be a counterpart. There is no known source associated with this flow, nor with MHO1382, which ends in a bright, knotty arc of emission. The shape of the arc suggests it is driving towards the cluster, not away from it. MHO1382 is associated with an extended red-shifted flow, with no detectable driving source. MHO1391 contains multiple arcs and knots of H$_2$ emission, each of the H$_2$ clusters associated with a peak of red-shifted CO emission - four in total. At the southern end of MHO1391, there is a compact blue-shifted emission region that overlaps the penultimate red-shifted region. However, there is no source that is obviously driving these flows. There are no detected sources at the southern end of MHO1391, although there is a prestellar core, DMM8 to the north, and seven {\it Spitzer}-identified protostars to the north-east, near DMM1. DMM1, located within 8$\arcsec$ of 5 {\it Spitzer}-identified protostars, is also coincident with MHO1388.

MHO1392 is part of the dominant region of red-shifted emission that extends across $\sim$8\arcmin\ to the south-west of NGC\,2264~D. It is crossed by a double-peaked region of blue-shifted emission, although there is no nearby source detected that could be driving this flow. MHO1393, MHO1395 and MHO 1394 are also coincident with this long ridge of red-shifted emission, although the different alignments of the knots, jets and bow shocks associated with these MHOs indicate that they may by arising in smaller flows that we are not able to distinguish from the larger scale flow. Further along the large red-shifted flow, are several more MHOs, MHO1397, MHO 1398 and MHO1399, each of which are coincident with peaks in the red-shifted flow. At the south-west tail of the large red flow is a compact bipolar flow, that has MHO3100 associated with the red-shifted lobe. There is no detected driving source for this flow, but it does lie outside the SCUBA and {\it Spitzer} observed areas. To the north of NGC\,2264~D, MHO1377 is coincident with a weak compact clump of blue-shifted emission, and a {\it Spitzer}-identified protostar, 12253 \citep{sung2009}.

\begin{table}
\begin{tabular}{@{}l@{}rrrrrr}
\hline
Source&\multicolumn{2}{c}{H$_2$}&\multicolumn{2}{c}{CO lobes}&\multicolumn{2}{c}{length ($\arcsec$)}\\
&Red&Blue &Red&Blue&Red&Blue\\
\hline
DMM7&&MHO1386&?&Y&&45\\
-&MHO1380&MHO1381&Y&Y&42&17\\
-&MHO1382&&Y&&66&\\
-&MHO3100&&Y&Y&35&52\\
12253$^a$&&MHO1377&M&Y&&19\\
\hline
\multicolumn{6}{l}{$^a$ Identification from \citet{sung2009}}\\
\end{tabular}
\caption{Driving source identifications for NGC\,2264~D flows.}
\label{tab-ngc2264-d} 
\end{table}

\label{lastpage}

\end{document}